\documentclass[12pt]{article}
\usepackage[T1]{fontenc}
\usepackage{amsfonts}
\usepackage{latexsym}
\usepackage{dsfont}
\usepackage{amsmath}
\usepackage{amssymb}
\usepackage{amssymb}
\usepackage{amsthm}
\usepackage{mathrsfs}
\usepackage[nosort]{cite}
\usepackage{setspace}
\usepackage{color}
\usepackage{graphicx}
\usepackage{accents}
\usepackage[font=footnotesize,labelfont=bf]{caption}
\usepackage[utf8]{inputenc}
\usepackage{geometry}
\usepackage[toc,page]{appendix}

\makeatletter
\newcommand{\labeltext}[2]{%
  \@bsphack
  \MakeLinkTarget*{#1}%
  \def\@currentlabel{#1}{\label{#2}}%
  \@esphack
}
\makeatother

\usepackage{tikz}
\usetikzlibrary{arrows.meta,backgrounds,calc,fit,positioning,scopes,shadows}

\makeatletter
\def\tikzsavelastnodename#1{\let#1=\tikz@last@fig@name}
\makeatother

\usepackage[hidelinks]{hyperref}
\hypersetup{
colorlinks=false,
citecolor= blue,
linkcolor= blue,
urlcolor= blue,
breaklinks=true
}
\usepackage{cleveref}
\crefformat{section}{\S#2#1#3} % see manual of cleveref, section 8.2.1
\crefformat{subsection}{\S#2#1#3}
\crefformat{subsubsection}{\S#2#1#3}

\def\appendix#1{\addtocounter{section}{1}\setcounter{equation}{0}
\renewcommand{\thesection}{\Alph{section}}
\section*{Appendix \thesection\protect\indent \parbox[t]{11.15cm}{#1}}
\addcontentsline{toc}{section}{Appendix \thesection\ \ \ #1}}

\allowdisplaybreaks
\numberwithin{equation}{section}

\makeatletter
 \let\old@startsection=\@startsection
 \let\oldl@section=\l@section
 \renewcommand{\@startsection}[6]{\old@startsection{#1}{#2}{#3}{#4}{#5}{#6\mathversion{bold}}}
 \renewcommand{\l@section}[2]{\oldl@section{\mathversion{bold}#1}{#2}}
\makeatother

\def\dd{\text{d}}

\DeclareMathOperator{\Tr}{Tr}
\DeclareMathOperator{\Pu}{P}

\begin{document}

%%%%%%%%%%%%% TITLE %%%%%%%%%%%%%%%%%%%%%%%%%

\begin{titlepage}
\vspace*{-1.0cm}

\begin{center}
%\today

\hfill {\footnotesize ZMP-HH/24-4\\
\hspace{13.21cm}TCDMATH-24-01}

\vspace{2.0cm}

{\LARGE  {\fontfamily{lmodern}\selectfont \bf Constructing Non-Relativistic AdS$_5$/CFT$_4$ \\
\vspace{2mm}Holography}} \\[.2cm]

\vskip 1.5cm
\textsc{Andrea Fontanella$^{\mathfrak{Re}}$ \footnotesize and \normalsize Juan Miguel Nieto Garc\'ia$^{\mathfrak{Im}}$}\\
\vskip 1.2cm

\begin{small}
{}$^{\mathfrak{Re}}$ \textit{School of Mathematics $\&$ Hamilton Mathematics Institute, \\
Trinity College Dublin, Ireland}\\
\vspace{1mm}
\href{mailto:andrea.fontanella[at]tcd.ie}{\texttt{andrea.fontanella[at]tcd.ie}}

\vspace{5mm}
{}$^{\mathfrak{Im}}$ \textit{II. Institut für Theoretische Physik, Universität Hamburg,\\
Luruper Chaussee 149, 22761 Hamburg, Germany} \\
\vspace{1mm}
\href{mailto:juan.miguel.nieto.garcia[at]desy.de}{\texttt{juan.miguel.nieto.garcia[at]desy.de}}

\end{small}

\end{center}

\vskip 1 cm
\begin{abstract}
\vskip1cm\noindent
We construct a new type of holographic correspondence between non-relativistic string theory in String Newton-Cartan AdS$_5\times$S$^5$ and Galilean Yang-Mills supplemented with 5 adjoint interacting scalar fields living on the 3+1 dimensional Penrose conformal boundary. In our derivation, we start with Maldacena's setting of a stack of $N$ coincident D3-branes and we show that the near-horizon/decoupling limit commutes with the non-relativistic limit, giving a unique answer both from the gravity and gauge theory perspectives. As a first evidence, we systematically compute the Killing vectors on the string theory side, and find that they match the symmetries of the dual gauge theory.

\end{abstract}

\end{titlepage}

\tableofcontents
\vspace{5mm}
\hrule

%%%%%%%%%%%%%%%%%%% BODY %%%%%%%%%%%%%%%%%%

\setcounter{section}{0}
\setcounter{footnote}{0}

\section*{Introduction}

\textcolor{red}{\underline{Correction}: In this article one should replace GED with GYM in every holographic statement, see the \ref{NoteAdded} at the end of section~\ref{sec:symm_gauge}}

%%%%%%% AdS/CFT a la Maldacena %%%%%%%
The holographic principle states the equivalence between two, a priori, very different theories: a theory of gravity and a theory of gauge interactions only. Its first concrete realisation was presented by Maldacena \cite{Maldacena:1997re} in the context of string theory, and later refined in \cite{Gubser:1998bc, Witten:1998qj, Maldacena:1998im}. The argument proposed by Maldacena was to consider Type IIB string theory in flat spacetime and add a stack of D3-branes in the bulk. This system admits two descriptions, one valid at small string coupling that describes the interaction of closed and open strings ending on the D3-brane, and the other description valid at large string coupling, where only closed strings propagate on the spacetime whose curvature is given by the presence of black D3-branes. The main observation is that both descriptions at low energies admit a decoupling limit. In the small string coupling description, the open and closed strings decouple at low energies and the system describes supergravity in 10d flat spacetime plus $\mathcal{N}=4$ Super Yang-Mills (SYM) theory in 4d flat spacetime. In the large string coupling regime, the spacetime geometry separates into two regions that decouple from each other at low energies: AdS$_5\times$S$^5$ near the horizon of the stack of D3-branes and flat spacetime far away from them. The core idea of holography is to identify these two descriptions at low energies, leading to the equivalence between  $\mathcal{N}=4$ SYM and fluctuations around the AdS$_5\times$S$^5$ geometry described by supergravity. 

There are stronger versions of this holographic paradigm, which states that the correspondence should also hold at any energy regime (strong version) and even at any values of the rank of the gauge group $N$, or equivalently, at any values of the string coupling (strongest version). The strong version of the AdS$_5$/CFT$_4$ correspondence has been quantitatively tested thanks to the integrability of both theories \cite{Bena:2003wd, Minahan:2002ve}, see \cite{Beisert:2010jr} for a review on the topic. A formal proof of the strongest version is still missing.           

In this paper, we study in detail how the non-relativistic limit can be incorporated into Maldacena's construction of the AdS$_5$/CFT$_4$ correspondence. Taking a limit, in general, is not a harmless operation, and just because two theories are equivalent before taking the limit, there is no guarantee that the equivalence will hold after the limit has been taken. Much of the troubles come from the fact the limit is in general a non-invertible operation and, moreover, the order in which we apply them may change the final result. For instance, flat space holography does not immediately follow from the AdS$_5$/CFT$_4$ correspondence, although the large radius limit takes AdS$_5\times$S$^5$ to Mink$_{10}$ (for recent reviews on flat space holography, see e.g. \cite{Raclariu:2021zjz, Donnay:2023mrd, Pasterski:2021raf}). For the same reason, a priori it is not granted that non-relativistic holography follows from the usual relativistic AdS/CFT correspondence. Similarly to flat space holography, non-relativistic holography is intrinsically ``non-AdS'' and in particular it is non-Lorentzian, since the spacetime geometry is no longer a Lorentzian manifold after taking the non-relativistic limit. Therefore, there are good reasons to study non-relativistic holography from the point of view of exploring how general the holographic principle is.

%%%%%%%% Non-relativistic string theory %%%%%%%%

The non-relativistic limit of string theory was first studied in the context of strings in flat spacetime \cite{Gomis:2000bd, Danielsson:2000gi} and later it was also applied to AdS$_5\times$S$^5$ \cite{Gomis:2005pg}. The initial idea was to rescale coordinates of the target space by a parameter $c$, and then take $c$ to infinity. In this process, the world-sheet remains relativistic.\footnote{One can make the world-sheet non-Lorentzian too by taking a limit on the intrinsic speed of light $\tilde{c}$ associated with the world-sheet geometry. The resulting theory has been dubbed Spin Matrix Theory \cite{Harmark:2017rpg, Harmark:2018cdl, Harmark:2019upf, Harmark:2020vll, Bidussi:2023rfs}, and it has the promising feature of capturing the behaviour of the relativistic AdS$_5$/CFT$_4$ correspondence at small `t Hooft coupling, but at finite $N$.} This is not a straightforward procedure, as one always find that some components of the metric diverge in the limit. To cancel the divergence, one needs to couple the string to a critical closed Kalb-Ramond B-field before taking the limit. The resulting spacetime geometry is non-Lorentzian, and is called ``String Newton-Cartan'' (SNC) geometry \cite{Andringa:2012uz, Bergshoeff:2018yvt, Bergshoeff:2019pij}. As the string world-sheet remains relativistic, one is still able to compute the beta function and show that the Weyl anomalies of the bosonic non-relativistic string action vanishes in $D=26$ \cite{Gomis:2019zyu,Gallegos:2019icg}.   

Different approaches to derive the non-relativistic string action have been studied. One of them is the so-called null reduction approach \cite{Harmark:2017rpg, Harmark:2018cdl, Harmark:2019upf}, which consists of taking a relativistic string action with a non-compact null isometry in target space and T-dualise it along this null direction while keeping the momentum fixed. The result is a non-relativistic action equivalent to the one obtained by the limit procedure. Another method to obtain the non-relativistic string action is via the expansion method \cite{Hartong:2021ekg, Hartong:2022dsx}, which is based on expanding the action in a large parameter $c$ and studying the equations of motion originating from the action at various orders in $c$. This method relies on the Lie algebra expansion, as a way to truncate the action at a certain higher order in $c$. For its application to the AdS$_5\times$S$^5$ string action, see \cite{Fontanella:2020eje}.

Non-relativistic string theory in SNC AdS$_5\times$S$^5$ has also been studied from the integrability point of view. Common techniques from Lorentzian theories cannot be borrowed and applied directly in this new context, as they need to be adapted accordingly to the non-Lorentzian nature of the theory. A coset construction of the non-relativistic action, as the analog of the Metsaev-Tseytlin for its relativistic counterpart, was given in \cite{Fontanella:2022fjd, Fontanella:2022pbm}. This way of rewriting the action found an immediate application in terms of finding a Lax connection \cite{Fontanella:2022fjd}. The spectral curve associated to it has also been studied in \cite{Fontanella:2022wfj}, which was found to be trivial due to the non semi-simplicity of the non-relativistic algebra. A generalisation of the spectral curve to the non-Lorentzian case was also proposed. Classical string solutions of the non-relativistic theory and the semiclassical expansion of the action were studied in \cite{Fontanella:2021btt, Fontanella:2023men, Fontanella:2021hcb}. For a recent review on aspects of non-relativistic string theory, see \cite{Oling:2022fft}. 

%%%%%%%%% Non-relativistic QFTs %%%%%%%%%%%%%%%%
On the gauge theory side, one can study how quantum field theories (QFTs) behave in the non-relativistic limit. Non-relativistic gauge theories are relevant for condensed matter systems, where Lorentz invariance of the microscopic description is not necessarily preserved, see e.g. cold atoms \cite{Nishida:2010tm}, fermions at unitarity \cite{Son:2008ye}, quantum Hall effect \cite{Geracie:2014nka}, strange metallic phases \cite{Hartnoll:2009ns} and quantum mechanical problems, like the Efimov effect \cite{Bedaque:1998km}. One of the simplest examples of non-relativistic QFTs is Galilean Electrodynamics (GED) \cite{Santos:2004pq}. This theory was proposed as a Lagrangian description of Maxwell's equations, written in the Galilei invariant formalism. This theory can be derived by the non-relativistic limit of a Maxwell 1-form potential $A_{\mu}$ and a real scalar field $\phi$ \cite{Bergshoeff:2015sic}, and its symmetries have been analysed in \cite{Festuccia:2016caf, Bagchi:2022twx, GED_symmetries}. A supersymmetric version of GED in 2+1 dimensions was given in \cite{Baiguera:2022cbp}, via null reduction. In addition to GED, also Galilean Yang-Mills theories have been studied in different aspects, see \cite{Bagchi:2022twx, Bagchi:2015qcw, Lambert:2019nti, Islam:2023iju}. Another example of supersymmetric and Galilean invariant QFT is the Wess-Zumino model found in \cite{Auzzi:2019kdd}. For a recent review on modern techniques of non-relativistic QFTs, see \cite{Baiguera:2023fus}.

%%%%%%%%%% Content of this paper %%%%%%%

In this paper, we propose a holographic duality between non-relativistic string theory and non-relativistic QFT. More precisely, we propose that non-relativistic string theory in SNC AdS$_5\times$S$^5$ is holographically dual to Galilean Electrodynamics with 5 additional uncharged massless free scalar fields. To our knowledge, this is the first holographic duality involving a non-relativistic theory of strings with relativistic world-sheet.\footnote{Once static gauge has been fixed and after expanding the embedding coordinates about the static string solution, the non-relativistic string theory action in SNC AdS$_5\times$S$^5$ becomes one of the free fields in AdS$_2$. In this setting, its holographic correspondence with a conformal quantum mechanics was studied in \cite{Sakaguchi:2007ba}. However, this type of holographic duality can be seen as the leading order in the coupling of the relativistic AdS$_2$/CFT$_1$ correspondence found in \cite{Giombi:2017cqn}, which in turns is induced from the AdS$_5$/CFT$_4$ correspondence. We remark that the holographic correspondence presented in this article, instead, comes from a brane construction, involving theories defined on a non-Lorentzian spacetime, and for these reasons is different.} While this work was in preparation, a similar idea was applied in the context of the M2-brane in M-theory \cite{Lambert:2024uue}. Holography involving non-relativistic string theory with non-relativistic world-sheet has been discussed in \cite{Harmark:2006ta,Harmark:2016cjq}. 
Other works on non-relativistic holography, although not in the context of string theory, are \cite{Taylor:2008tg, Taylor:2015glc,CaldeiraCosta:2010ztk, Guica:2010sw, Hartong:2013cba, Christensen:2013lma, Christensen:2013rfa, Hartong:2014oma, Hartong:2014pma, Dobrev:2013kha}.

\begin{figure}[t!]
    \centering
    \scalebox{0.9}{\begin{tikzpicture}[> = {Triangle[]},atomic/.style = {draw, thick, fill=white, font=\footnotesize,minimum size=1.5cm, drop shadow,append after command= {\pgfextra{\tikzsavelastnodename\tikzsavednodename}},#1}]
%---
\node[atomic,align=center]                         (D3)     {Stack of \\[1mm]
D3 branes};

\node[atomic,align=center,right=3cm of D3]                         (NHL)     {Strings in \\[1mm]
 \, AdS$_5 \times$ S$^5$ \, };

\node[atomic,align=center,below=1cm of D3]                         (NRD3)     {Stack of \\[1mm]
NR D3 branes};

\node[atomic,align=center,below=1cm of NHL]                         (NRNHL)     {NR Strings in \\[1mm] 
SNC AdS$_5 \times$ S$^5$};

\node[atomic,align=center,right=3cm of NHL]                         (N4SYM)     {$\mathcal{N}=4$ SYM};

\node[atomic,align=center,below=1cm of N4SYM]                         (NRN4SYM)     {GED\\
with scalars};

\node[atomic,align=center,right=3cm of N4SYM]                         (DBI)     {DBI action \\
in Mink$_{10}$};

\node[atomic,align=center,below=1cm of DBI]                         (NRDBI)     {DBI action in\\
SNC Mink$_{10}$};

\draw[->] (D3) -- node [font=\footnotesize,midway,above=0.2cm,align=center ]{$\alpha ' \rightarrow 0$} (NHL);
\draw[->] (NHL) -- node [font=\footnotesize,midway,right=0.2cm,align=center ]{$c\rightarrow \infty$} (NRNHL);
\draw[->] (NRD3) -- node [font=\footnotesize,midway,above=0.2cm,align=center ]{$\alpha ' \rightarrow 0$} (NRNHL);
\draw[->] (D3) -- node [font=\footnotesize,midway,right=0.2cm,align=center ]{$c\rightarrow \infty$} (NRD3);
\draw [line width=1pt, double distance=2pt,
arrows = {Latex[length=0pt 3 0]-Latex[length=0pt 3 0]}]  (NHL) -- node [font=\footnotesize,midway,above=0.2cm,align=center ]{Duality} (N4SYM);
\draw [line width=1pt, double distance=2pt,
arrows = {Latex[length=0pt 3 0]-Latex[length=0pt 3 0]}] (NRNHL) -- node [font=\footnotesize,midway,above=0.2cm,align=center ]{New duality} (NRN4SYM);
\draw[->] (N4SYM) -- node [font=\footnotesize,midway,right=0.2cm,align=center ]{$c\rightarrow \infty$} (NRN4SYM);
\draw[<-] (N4SYM) -- node [font=\footnotesize,midway,above=0.2cm,align=center ]{$\alpha ' \rightarrow 0$} (DBI);
\draw[<-] (NRN4SYM) -- node [font=\footnotesize,midway,above=0.2cm,align=center ]{$\alpha ' \rightarrow 0$} (NRDBI);
\draw[->] (DBI) -- node [font=\footnotesize,midway,right=0.2cm,align=center ]{$c\rightarrow \infty$} (NRDBI);
\end{tikzpicture}}
\vspace{1mm}

  \caption{The top line of the diagram is the well-known relativistic AdS$_5$/CFT$_4$ correspondence. The bottom line is the new correspondence between non-relativistic theories found in this paper. In the construction, the near-horizon/decoupling limit $\alpha'\to 0$ commutes with the non-relativistic limit $c\to \infty$. }
    \label{fig:NR_holography}
\end{figure}
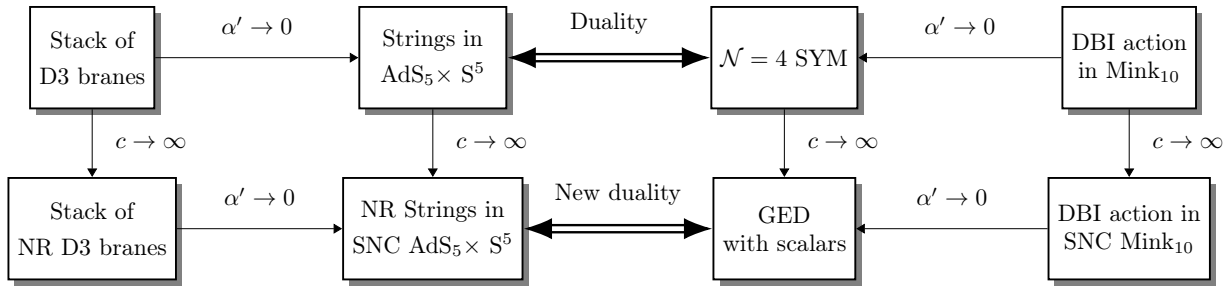

The plan of this paper is the following. In section \ref{sec:NR_decoupling} we study in detail the compatibility between the decoupling and the non-relativistic limits. In particular, in section \ref{sec:gravity_persp} we start with a stack of $N$ black D3-branes in Type IIB string theory, and we apply the near-horizon limit followed by the non-relativistic one, and then the other way around. In both cases we get the same final result, namely the SNC AdS$_5\times$S$^5$ geometry. Furthermore, we show that the non-relativistic metric for the stack of black D3-branes retains a notion of horizon. We conclude this section by showing that SNC AdS$_5\times$S$^5$ admits a Penrose conformal boundary, given by NC Mink$_4$, which will turn out to be the spacetime where the holographic gauge theory is defined. 

In section \ref{sec:gauge_persp} we consider the Dirac-Born-Infeld (DBI) action in Mink$_{10}$ for a stack of D3-branes. We take the decoupling limit followed by the non-relativistic limit, and then the other way around. Also, from this gauge perspective, we show that we obtain the same final result, given by GED in 3+1 dimensions supplemented by 5 uncharged massless free scalars. We end this section by showing that there is a second possibility of taking the non-relativistic limit, which gives a Galilean Yang-Mills theory in 3+1 dimensions minimally coupled to 5 charged massless scalars with a non-trivial potential. In this case, the gauge group is not abelian as it happens to be in the first limit, but retains its $U(N)$ structure. However, we do not provide a holographic description for this gauge theory. Although the non-relativistic limit of the DBI action has already been studied in \cite{Gomis:2020fui}, this paper only considers the case of a single $(d-2)$-brane. Our starting point DBI action describes a stack of $N$ coincident D3-branes, keeping track of the non-abelian nature of the system.

In section \ref{sec:global_symm} we support our holographic claim by providing a first test based on matching the symmetries of the two theories. In particular, in section \ref{sec:symm_string} we systematically compute the non-relativistic Killing vectors that leave the string action on SNC AdS$_5\times$S$^5$ invariant. We find that this isometry algebra is infinite dimensional, and contains an $\mathfrak{sl}(2,\mathbb{R})$ subalgebra. In section \ref{sec:symm_gauge} we match the string theory symmetries with the symmetries of GED in 3+1 dimensions, with 5 uncharged massless scalar fields, which were computed in \cite{GED_symmetries}. 
The findings of this research are summarised in Figure \ref{fig:NR_holography}. For a concise letter version of the results of this article, we refer to \cite{Fontanella:2024kyl}.

\section{The non-relativistic and decoupling limits}
\label{sec:NR_decoupling}

In this section, we shall discuss the compatibility of the decoupling and non-relativistic limits in both gravity and gauge perspectives.
The non-relativistic limit that we consider is ``stringy'', in the sense that two coordinates (a time-like and a space-like) are rescaled with a parameter $c$. This is the generalisation to the string of the usual ``particle'' non-relativistic limit, where only a time-like coordinate is rescaled with $c$.\footnote{The parameter $c$ is always taken to be dimensionless. Even in the particle non-relativistic limit, where $c$ has physical meaning of speed of light, one should not take the limit directly on $c$, but should rescale $c \to \omega c$, with $\omega$ a new dimensionless parameter, and then take $\omega \to \infty$.} 
The stringy non-relativistic limit changes the intrinsic Lorentzian geometry of spacetime, which becomes a codimension 2 foliation described by a String Newton-Cartan (SNC) geometry characterised by two metrics, a longitudinal one $\tau_{\mu\nu}$ and a transverse one $h_{\mu\nu}$. The reason to focus on the stringy instead of the particle non-relativistic limit is that the latter one has been shown to produce a trivial theory where strings do not vibrate \cite{Batlle:2016iel}, something that does not happen in the stringy limit.

\subsection{The gravity perspective}
\label{sec:gravity_persp}

We start by writing the spacetime metric given by a stack of $N$ black D3-branes,\footnote{Our conventions and a summary of indices used in this paper are given in Appendix \ref{app:Conventions}.}   
\begin{eqnarray}
\label{D3_metric}
    \dd s^2_{\text{D3-brane}} &=& \frac{1}{\sqrt{f(r)}} \left[ -(\dd \tilde{t})^2 + \dd \tilde{x}^i \dd \tilde{x}_i   \right]  + \sqrt{f(r)} (\dd r^2 + r^2 \dd \Omega^2_5 ) \, , \\
\notag
f (r) &=& 1 + \frac{4 \pi  g_s N \alpha'^2 }{r^4} \, ,
\end{eqnarray}
where $(\tilde{t}, \tilde{x}^i)$, $i=1,2,3$, are coordinates along the world-volume of the D3-brane, and $\dd \Omega_5^2$ is the metric of the unit 5-sphere. We describe the 5-sphere metric in terms of Cartesian coordinates $(\phi, y^m)$, where $m=1,..., 4$ and $y^2 \equiv y^m y^n \delta_{mn}$, given by 
\begin{eqnarray}
   \dd \Omega_5^2 = \left(\frac{1-\frac{y^2}{4}}{1+\frac{y^2}{4}}\right) \dd \phi^2 + \frac{\dd y^m \dd y^n \delta_{mn}}{\left( 1+\frac{y^2}{4}\right)^2} \, . 
\end{eqnarray}
To make contact with AdS$_5$ in the Poincar\'e patch, we perform the following change of coordinate: 
\begin{eqnarray}
\label{z_coord}
    r = \frac{\alpha'}{z} \, ,
\end{eqnarray}
such that the metric (\ref{D3_metric}) becomes 
\begin{eqnarray}
\label{D3_metric_z}
    \dd s^2_{\text{D3-brane}} &=& \frac{1}{\sqrt{f(z)}} \left[ -(\dd \tilde{t})^2 + \dd \tilde{x}^i \dd \tilde{x}_i   \right]  + \alpha'^2 \sqrt{f(z)} \left(\frac{\dd z^2}{z^4} + \frac{1}{z^2} \dd \Omega^2_5 \right) \, , \\
\notag
f (z) &=& 1 + \frac{4 \pi  g_s N}{\alpha'^2} z^4 \, .
\end{eqnarray}
Now we want to show that taking the near-horizon limit and the non-relativistic limit on (\ref{D3_metric_z}) commutes.

\paragraph{Near-horizon first, non-relativistic second}

We take the metric (\ref{D3_metric_z}) and we rescale the coordinates along the world-volume of the D3-brane as follows:
\begin{eqnarray}
\label{world_volume_rescaling}
    \tilde{t} = \sqrt{4 \pi g_s N} t \, , \qquad\qquad
    \tilde{x}^i = \sqrt{4 \pi g_s N} x^i \, . 
\end{eqnarray}
Then we take the near-horizon limit $\alpha' \to 0$. This will give us 
\begin{eqnarray}
\label{AdS5xS5_metric}
\dd s^2_{\text{AdS}_5\times\text{S}^5} = \sqrt{4 \pi  g_s N} \alpha' \left[\frac{1}{z^2} \left( -\dd t^2 +  \dd z^2 + \dd x^i \dd x^j \delta_{ij} \right)  + \dd \Omega_5^2 \right]\, ,
\end{eqnarray}
which is the AdS$_5\times$S$^5$ metric, with the same radius $R$ for both AdS$_5$ and S$^5$, given by  
\begin{eqnarray}
\label{radius_def}
    R^2 \equiv \sqrt{4 \pi  g_s N} \alpha' \, . 
\end{eqnarray}

The next step is to take a stringy non-relativistic limit. The limit we consider rescales the coordinates $(t, z)$, which we call longitudinal, differently from the remaining coordinates, which we denote as transverse. This limit is implemented by rescaling the radius $R$ and the transverse coordinates by a parameter $c$ as follows: 
\begin{eqnarray}
\label{NR_rescaling}
    R \to c \, R \, , \qquad
    x^i \to \frac{x^i}{c} \, , \qquad 
    \phi \to \frac{\phi}{c} \, , \qquad
    y^m \to \frac{y^m}{c} \,  .
\end{eqnarray}
In the $c\to \infty$ limit, we find that the metric becomes the sum of two terms scaling with different powers of $c^2$, namely   
\begin{subequations}
\label{SNC_AdS5xS5_metric}
\begin{align}
    \dd s^2_{\text{SNC AdS}_5\times\text{S}^5} &= (c^2 \tau_{\mu\nu} + h_{\mu\nu}) \dd X^{\mu} \dd X^{\nu}\,  , \\
    \tau_{\mu\nu} \dd X^{\mu} \dd X^{\nu} &= \frac{R^2}{z^2} \left( -\dd t^2 + \dd z^2 \right) \, , \\
    h_{\mu\nu} \dd X^{\mu} \dd X^{\nu} &= \frac{R^2}{z^2} \dd x^i \dd x^j \delta_{ij} + R^2 \dd x^{i'} \dd x^{j'} \delta_{i' j'} \, , 
\end{align}
\end{subequations}
where we denoted collectively the flat coordinates originating from the 5-sphere, $x^{i'} \equiv (\phi, y^m)$, $i'=5,...,9$, and $X^0 \equiv t,  X^i \equiv x^i , X^4 \equiv z , X^{i'} \equiv x^{i'}$
The geometry so obtained is called ``String Newton-Cartan geometry'' of AdS$_5\times$S$^5$, and it is defined by the tensors $\tau_{\mu\nu}, h_{\mu\nu}$. The metric $\tau_{\mu\nu}$ describes the AdS$_2$ spacetime, whereas $h_{\mu\nu}$ describes a warped Euclidean space, $w(z) \mathbb{R}^3 \times \mathbb{R}^5$, where $w(z) = z^{-2}$ is a warped factor for the transverse AdS$_5$ directions which have flattened out in the limit. The 5-sphere coordinates have also flattened out in the non-relativistic limit, and we have collectively denoted its coordinates by $x^{i'} \equiv (\phi , y^m)$. The geometry is invariant under the local String Newton-Cartan transformations \cite{Bergshoeff:2019pij}, under which only $\tau_{\mu\nu}$ is a metric, as it does not change signature. On the other hand, $h_{\mu\nu}$ can change signature, and therefore it is just a tensor. In String Newton-Cartan geometry, $\tau_{\mu\nu}$ and $h^{\mu\nu}$ have the meaning of metrics, whereas $\tau^{\mu\nu}$ and $h_{\mu\nu}$ are tensors. 

Studying closed strings in the background geometry (\ref{SNC_AdS5xS5_metric}) requires coupling the string to a critical closed Kalb-Ramond B-field $B_{\mu\nu} = c^2 \tau_{\mu}{}^A \tau_{\nu}{}^B \varepsilon_{AB}$, where $\tau_{\mu}{}^A$ is the SNC vielbein associated with the $\tau_{\mu\nu}$ metric. 
After applying a Hubbard-Stratonovich transformation to the $c^2$ term coming from the metric and the critical B-field, one obtains a finite action, given by the non-relativistic string action in SNC AdS$_5\times$S$^5$ \cite{Gomis:2005pg}.

\paragraph{Non-relativistic first, near-horizon second}

Now we want to take the limits in the opposite order than what we did previously and show that we arrive at the same final result. We start by taking the metric (\ref{D3_metric_z}), we redefine $\tilde{t}$ and $\tilde{x}^i$ as given in (\ref{world_volume_rescaling}) and we rewrite any $\sqrt{4 \pi  g_s N}$ factor in terms of the radius as given in (\ref{radius_def}). Then we rescale coordinates as required by the non-relativistic limit. This amounts to taking the rescaling (\ref{NR_rescaling}) supplemented by a rescaling of $\alpha'$. Concretely,\footnote{One can formally perform the rescaling $\alpha' \to c \, \alpha'$ also in (\ref{NR_rescaling}). However, in that case we already took $\alpha' \to 0$, so there is no $\alpha'$ parameter to act on apart from those appearing inside the definition of the radius $R$. For consistency, rescaling $R\to c R$ and $\alpha'\to c \alpha'$ implies that $g_s N$ rescales as $c^2 g_s N$. }
\begin{eqnarray}
\label{NR_rescaling_alpha_p}
    \alpha' \to c \, \alpha' \, , \qquad
    R \to c \, R \, , \qquad
    x^i \to \frac{x^i}{c} \, , \qquad 
    \phi \to \frac{\phi}{c} \, , \qquad
    y^m \to \frac{y^m}{c} \,  .
\end{eqnarray}
Then we take the $c\to \infty$ limit, and we get the metric of the non-relativistic stack of D3-branes,    
\begin{subequations}
\label{NR_D3_metric}
\begin{align}
    \dd s^2_{\text{NR D3-brane}} &= \left(c^2 \tau_{\mu\nu} + h_{\mu\nu} \right) \dd X^{\mu} \dd X^{\nu}  \,  , \\
    \tau_{\mu\nu} \dd X^{\mu} \dd X^{\nu} &= -\frac{R^4}{\alpha^{\prime 2}} \frac{1}{\sqrt{f(z)}} \dd t^2 +  \alpha^{\prime 2}\sqrt{f(z)} \, \frac{\dd z^2}{z^4} \, , \\
    h_{\mu\nu} \dd X^{\mu} \dd X^{\nu} &= \frac{R^4}{\alpha^{\prime 2}} \frac{1}{\sqrt{f(z)}} \dd x^i \dd x^j \delta_{ij} + \alpha^{\prime 2}\sqrt{f(z)} \frac{1}{z^2} \dd x^{i'} \dd x^{j'} \delta_{i' j'}  \, , \\
    f(z)&= 1 + \frac{R^4}{\alpha^{\prime 4}} z^4 \, . 
\end{align}
\end{subequations}
The reason for rescaling $\alpha'$ with $c$ in (\ref{NR_rescaling_alpha_p}) is that, in this way, the non-relativistic metric (\ref{NR_D3_metric}) still admits a near-horizon region that can be decoupled from the rest of the spacetime.\footnote{A different non-relativistic limit of the black D3-brane solution was studied in \cite{Avila:2023aey} under the name of ``transverse black brane''. This solution is flat almost everywhere (so it is not interesting for our purposes), and it satisfies the non-relativistic NS-NS supergravity equations of motion of \cite{Gomis:2019zyu,Gallegos:2019icg}. To check that our proposed non-relativistic black D3-brane (\ref{NR_D3_metric}) satisfies the non-relativistic supergravity equations of motion requires including the RR forms to source the curved geometry, which we leave for future work.} On the other hand, if we did not rescale $\alpha'$ in (\ref{NR_rescaling_alpha_p}),  taking the $c \to \infty$ limit would immediately bring the relativistic metric of a stack of D3-branes (\ref{D3_metric_z}) to the metric of SNC AdS$_5\times$S$^5$ (\ref{SNC_AdS5xS5_metric}). This result then would not be suitable for the purpose of formulating holography, as the SNC AdS$_5\times$S$^5$ spacetime does not exhibit a near-horizon region.

Finally, we take the near-horizon limit  on the non-relativistic stack of D3-branes (\ref{NR_D3_metric}). This is obtained by taking the $\alpha' \to 0$ limit\footnote{From the perspective of the old $\alpha'$ variable in (\ref{NR_rescaling_alpha_p}), we want that $c\, \alpha' \to 0$. This means that we demand the new $\alpha'$ variable to go to zero faster than how $c$ goes to infinity. Notice that this fits with the other order of taking the limits, where we take $\alpha'\to 0$ first, so in that case $c \alpha' \to 0$ is a natural consequence of the construction.}, and it gives us precisely the SNC AdS$_5\times$S$^5$ metric (\ref{SNC_AdS5xS5_metric}). Therefore, this proves the near-horizon limit commutes with the non-relativistic one.

\subsubsection{The conformal boundary}
\label{sec:conf_boundary}

In this section we find the conformal boundary geometry of SNC AdS$_5\times$S$^5$ via the Penrose procedure. As we shall see in section \ref{sec:gauge_persp}, this is precisely the manifold where the non-relativistic holographic dual gauge theory lives on. 
Following Penrose \cite{Penrose}, the asymptotic behaviour of a given manifold $\mathcal{M}$ with metric $g$, namely the study of its points at infinity, is a problem that can be systematically formulated by making use of a conformal rescaling. Concretely, one can map a pseudo-Riemannian space $(\mathcal{M}, g)$ into another $(\tilde{\mathcal{M}}, \tilde{g})$, via the conformal transformation
\begin{eqnarray}
    \tilde{g}_{\mu \nu} = \Omega^2 g_{\mu \nu} \, . 
\end{eqnarray}
If $\Omega$ is chosen accordingly, any point at infinite distance, when measured with respect to $g$, can be taken at finite distance when measured with respect to $\tilde{g}$. This requires $\Omega$ to vanish at infinity. The set of points satisfying $\Omega = 0$ defines the \emph{conformal boundary} of $\mathcal{M}$.\footnote{In Penrose's formalism, one should demand an additional condition, namely that the gradient of $\Omega$ is non-vanishing on the conformal boundary, i.e. $\dd \Omega |_{\Omega = 0} \neq 0$. However, as pointed out by \cite{Norman} (see also the second footnote in page 182 of \cite{Penrose}), this condition is necessarily violated for some cosmological models and therefore can somewhat be dropped.}

The Penrose formalism can also be applied to non-Lorentzian manifolds. Thus, we will apply it to compute the conformal boundary of SNC AdS$_5\times$S$^5$. By adapting Poincar\'e coordinates to the AdS space, the SNC vielbeine of AdS$_5\times$S$^5$ are  
\begin{eqnarray}
    \tau_{\mu}{}^A = z^{-1} \delta_{\mu}^A \, , \qquad\qquad
    e_{\mu}{}^a = z^{-1} \delta_{\mu}^a \, , \qquad\qquad
    e_{\mu}{}^{a'} = \delta_{\mu}^{a'} \, ,
\end{eqnarray}
where $A=0,4$, $a=1,2,3$ and $a'=5,...9$ are the flat indices.
In these coordinates, we immediately see that the vielbein components diverge at $z=0$. However, this divergence is removed by rescaling each SNC vielbein via the following conformal factor, 
\begin{eqnarray}
    \tilde{\tau}_{\mu}{}^A = \Omega \, \tau_{\mu}{}^A \, , \qquad
    \tilde{e}_{\mu}{}^a = \Omega \, e_{\mu}{}^a \, , \qquad
    \tilde{e}_{\mu}{}^{a'} = \Omega \, e_{\mu}{}^{a'} \, , \qquad\qquad
    \Omega = z \, ,
\end{eqnarray}
and the conformal boundary is given by the points that make $\Omega$ vanishing, i.e. $z = 0$. The geometry of the conformal boundary is obtained by plugging the condition $z=0$ inside the rescaled vielbeine $\tilde{\tau}_{\mu}{}^A, \tilde{e}_{\mu}{}^a, \tilde{e}_{\mu}{}^{a'}$. By doing this, the geometry changes from being String Newton-Cartan to Newton-Cartan only. In this case, the conformal boundary geometry is described by the metric $\tilde{\tau}_{\mu\nu}$ and tensor $\tilde{h}_{\mu\nu}$ given by 
\begin{eqnarray}
    \tilde{\tau}_{\mu\nu} \dd X^{\mu} \dd X^{\nu} = - \dd t^2 \, , \qquad\qquad
    \tilde{h} _{\mu\nu} \dd X^{\mu} \dd X^{\nu} = \dd x^i \dd x^j \delta_{ij} \, ,  \label{NCMinkowski}
\end{eqnarray}
namely the Newton-Cartan geometry of 4d Minkowski spacetime, NC Mink$_4$. Similarly to what is discussed in \cite{Hartong:2013cba}, we expect that the String Newton-Cartan local symmetries of the bulk geometry SNC AdS$_5\times$S$^5$ will induce in the conformal boundary NC Mink$_4$ the expected local symmetries of a genuine Newton-Cartan manifold, namely the Bargmann transformations.

\subsection{The gauge theory perspective}
\label{sec:gauge_persp}

Let us now consider the weak string coupling regime, which corresponds to the gauge theory perspective of the correspondence. In this regime, we have both open and closed strings propagating in flat space, where open strings end on the D3 branes. Such system is governed by the following action
\begin{eqnarray}
\label{S_initial}
    S = S_{\text{closed}} + S_{\text{open}} + S_{\text{interaction}} \, ,
\end{eqnarray}
where $S_{\text{closed}}$ describes the dynamics of closed strings, $S_{\text{open}}$ does the same for open strings, and $S_{\text{interaction}}$ describes the interaction between the two kinds. In the decoupling limit $\alpha' \to 0$, the closed strings sector gives us supergravity in 10d, whereas the open strings sector gives us $\mathcal{N}=4$ SYM. In this limit $S_{\text{interaction}} \sim \alpha^{\prime 2}\rightarrow 0$, meaning that open and closed strings do not interact with each other. As $S_{\text{interaction}}$ and $S_{\text{open}}$ come with the same power of $c$ in the non-relativistic limit, this decoupling is true regardless of the order of the limits.

Open strings attached to D$p$-branes are described by the DBI action at low energies. However, as we are interested in analysing the case of coincident D3 branes, we have to make use of the non-abelian generalisations described in \cite{Tseytlin:1997csa,Myers:1999ps},\footnote{Although the action (\ref{improvedDBI}) is not correct beyond fifth order in $\alpha '$, we will only need the leading order in $\alpha '$ for our argument. }
\begin{equation}
    S_{\text{open}}=-T_4 \int{\dd \sigma^4 \Tr_S \left[ e^{-\Phi} \sqrt{-\det (Q^p{}_q) \det \left( \Pu_c\left[ E_{\mu\nu} + E_{\mu r} (Q^{-1}-\delta)^r\null_\nu  \right]_{\mathtt{mn}} + 2\pi \alpha' F_\mathtt{mn}  \right)} \right] } \, , \label{improvedDBI}
\end{equation}
where $T_4$ is the brane tension, $E=g+B_2$ is the combination of the metric and the Kalb-Ramond field; $F_\mathtt{mn} = \partial_\mathtt{m} A_\mathtt{n} -\partial_\mathtt{n} A_\mathtt{m} +i [A_\mathtt{m} , A_\mathtt{n}]$ is the field strength of a gauge potential $A_\mathtt{m} = A_\mathtt{m}^I T^I$, where $T^I$ are generators of $U(N)$ that carry the Chan-Paton indices; $\Pu_c [E_{\mu\nu}]_\mathtt{mn}$ indicates the pullback of $E_{\mu\nu}$, with the usual derivatives replaced by covariant derivatives $D_\mathtt{m} X^{\mu}= \partial_\mathtt{m} X^{\mu} + i [A_\mathtt{m} , X^{\mu}]$; and $\Tr_S$ is the symmetrised trace over Chan-Paton indices, that is, the trace over the sum of all the ways we can permute the matrices
\begin{equation}
    \Tr_S (A_1 \dots A_n) = \Tr( A_1 \dots A_n + \text{perm.}) \, .
\end{equation}
Finally, the matrix $Q$ takes the form
\begin{equation}
    Q^p\null_q = \delta^p\null_q +\frac{i [X^p , X^r]}{2\pi \alpha '} E_{rq} \, ,
\end{equation}
where the indices $p$, $q$ and $r$ only run over the transverse coordinates to the D3-brane world-volume, $p,q,r = 4, ..., 9$.

\paragraph{Non-relativistic first, decoupling second}

Let us consider first the non-relativistic limit of the DBI action (\ref{improvedDBI}). For that, we will follow similar steps as in \cite{Gomis:2020fui}, although we implement the non-relativistic limit differently. 

We start by fixing our metric $g$ to be the flat Minkowski metric in 10 dimensions. Then we need to take a stringy non-relativistic limit, in analogy to what we did in the gravity perspective. To define such a limit, there are two ambiguities to resolve. First, we have to decide which directions will be longitudinal and which ones will be transverse (with respect to the non-relativistic limit). As time will always be longitudinal, we only need to declare whether the longitudinal spatial direction is a coordinate that will be spanning the world-volume once we will fix static gauge at a later stage, or whether it is a coordinate belonging to its transverse space. We choose it to be the second case, because in this way the final Lagrangian will be defined on a 3+1 Newton-Cartan spacetime that matches the conformal boundary structure we found in section \ref{sec:conf_boundary}. 

The second ambiguity corresponds to deciding if we either rescale the longitudinal coordinates with a factor $c$ or we rescale the transverse coordinates with a factor $1/c$. Here we will choose the latter, which we will justify a posteriori by a holographic matching of symmetries. Nevertheless, we will discuss the former rescaling in section~\ref{sec:alternative}.

Motivated by this discussion, we rescale the target space coordinates as
\begin{eqnarray}
\label{NR_rescaling_gauge_1}
    X^\mathtt{a} \to \frac{X^\mathtt{a}}{c}  \,  , \qquad X^{i'} \to \frac{X^{i'}}{c}  \,  .
\end{eqnarray}
In this and subsequent expressions, $\mathtt{a}$ and $\mathtt{b}$ run only over the spatial world-volume coordinates, i.e., $\sigma^\mathtt{m}=(\mathtt{t}, \sigma^\mathtt{a})$, and $i'$ runs from $5$ to $9$. We have chosen to perform the rescaling on these coordinates because $X^0$ is the only time-like coordinate available and $X^4$ is the first spatial coordinate that we do not fix later by imposing static gauge. Once this rescaling is applied, the pullback of the metric splits into 
\begin{equation}
    \Pu_c [g_{\mu \nu}]_{\mathtt{m}\mathtt{n}}= \left( \tau_\mu{}^A \tau_\nu{}^B \eta_{AB}  + \frac{1}{c^2} h_{\mu \nu} \right)  D_\mathtt{m} X^\mu D_\mathtt{n} X^\nu \, ,
\end{equation}
where $\tau_{\mu}{}^A = \delta^A_{\mu}$ and $h_{\mu\nu} = \delta_{\mu\nu}$.
In order to not have divergent terms when we compute the non-relativistic limit, we need to set all the components of $B_2$ to zero except for $B_{04}$, which we fix to 
\begin{equation}
    B_{\mu \nu}= \tau^A_\mu \tau^B_\nu \varepsilon_{AB} \, , \label{B_field_gauge}
\end{equation}
allowing us to write the pullback of $E$ as
\begin{equation}
    \Pu_c[E_{\mu \nu}]_{\mathtt{m}\mathtt{n}}= \left( -\bar{\tau}_\mu \tau_\nu + \frac{1}{c^2} h_{\mu \nu}  \right)  D_\mathtt{m} X^\mu D_\mathtt{n} X^\nu \, ,
\end{equation}
where $\bar{\tau}_\mu =\tau^0_\mu -\tau^4_\mu $, ${\tau}_\mu =\tau^0_\mu +\tau^4_\mu $.

Let us focus now on the matrix $Q$. After applying the transformation (\ref{NR_rescaling_gauge_1}), we can split it into three contributions depending on the power of $c$,
\begin{equation}
    Q^p\null_q =\delta^p\null_q +\frac{1}{c} \left( \delta^{p4} \sum_{r=5}^9 \frac{i [X^4 , X^r]}{2\pi \alpha '} E_{rq} + \frac{i [X^p , X^4]}{2\pi \alpha '} E_{4q}\right)    + \frac{1}{c^2} \sum_{r=5}^9 (1-\delta^{p4}) \frac{i [X^p , X^r]}{2\pi \alpha '} E_{rq} \, ,
\end{equation}
from which we obtain
\begin{align}
    &(Q^{-1} - \delta)^p\null_q =-\frac{1}{c} \left( \delta^{p4} \sum_{r=5}^9 \frac{i [X^4 , X^r]}{2\pi \alpha '} E_{rq} + \frac{i [X^p , X^4]}{2\pi \alpha '} E_{4q}\right)    - \frac{1}{c^2} \sum_{r=5}^9 (1-\delta^{p4}) \frac{i [X^p , X^r]}{2\pi \alpha '} E_{rq} \notag \\
    &+ \frac{1}{c^2} \left( \delta^{p4} \sum_{r,s=5}^9 \frac{i [X^4 , X^r]}{2\pi \alpha '} E_{rs} \frac{i [X^s , X^4]}{2\pi \alpha '} E_{4q} + \sum_{r=5}^9 \frac{i [X^p , X^4]}{2\pi \alpha '} E_{44} \frac{i [X^4 , X^r]}{2\pi \alpha '} E_{rq} \right) +\mathcal{O} \left( c^{-3} \right) \, .
\end{align}
Once we apply the pullback to this quantity multiplied by $E$ from the left, and keep terms only up to $c^{-2}$, we get
\begin{align}
    &\Pu_c[E_{\mu p} (Q^{-1} - \delta)^p\null_\nu]_{\mathtt{m}\mathtt{n}} =-\frac{2}{c^2} \sum_{q,r=5}^9 \frac{i [X^4 , X^r]}{2\pi \alpha '} \tau_0{}^A \tau_4{}^B \varepsilon_{AB} h_{rq}D_\mathtt{m} X^0 D_\mathtt{n} X^q\\
    &+ \frac{1}{c^2} \sum_{\mu,\nu} D_\mathtt{m} X^\mu E_{\mu4} \sum_{r,s=5}^9 \frac{i [X^4 , X^r]}{2\pi \alpha '} E_{rs} \frac{i [X^s , X^4]}{2\pi \alpha '} E_{4\nu} D_\mathtt{n} X^\nu +\mathcal{O} \left( c^{-3} \right) \equiv \frac{q_\mathtt{mn}}{c^2} +\mathcal{O} \left( c^{-3} \right)\, . \notag 
\end{align}
Substituting these expressions into (\ref{improvedDBI}) gives us
\begin{align}
    \mathcal{S}_{\text{NR DBI}} =& -T_4 \int{\dd \sigma^4 \Tr_S \left[ e^{-\Phi} \sqrt{ -\det (Q^p{}_q)} \right. } \notag \\
    \cdot & \left. \sqrt{\det \left[ \left( -\bar{\tau}_\mu \tau_\nu + \frac{1}{c^2} h_{\mu \nu} \right)  D_\mathtt{m} X^\mu D_\mathtt{n} X^\nu + 2\pi \alpha' F_\mathtt{mn}+ \frac{q_\mathtt{mn}}{c^2}   + \mathcal{O} \left( c^{-3} \right)   \right]} \right]  \, .
\end{align}
In order to have a consistent limit, we also need to rescale the gauge fields as follows:\footnote{By consistent limit, we mean that the decoupling and non-relativistic limits commute, leading to the same final action. This, combined with demanding that all the degrees of freedom appear at leading order in the expansion in $c$, determines the rescaling (\ref{NR_rescaling_gauge_2}).} 
\begin{equation}
    A_\mathtt{m} \to \frac{A_\mathtt{m} }{c^2} \, . \label{NR_rescaling_gauge_2}
\end{equation}
This transformation has important implications for the gauge group structure of our action, as it shifts all non-abelian structure constants to higher orders in $c$. At leading order, all covariant derivatives become partial derivatives and the commutator term in the field strength disappears. Effectively, this abelianises the gauge group from $U(N)$ to $U(1)^{N^2}$.\footnote{The rescaling (\ref{NR_rescaling_gauge_2}) resembles the weak-coupling limit, as both eliminate the non-abelian structure. As the 3+1 spacetime is non-compact, we do not have to deal with subtleties related to global structures, as in the compact case.}   

The action we have obtained can be written in a simpler fashion by performing similar transformations to the ones done in \cite{Bergshoeff:2019pij,Gomis:2020fui}. For that, we need to use the Sylvester determinant identity for rectangular matrices to transform the determinant over the indices $\mu$ and $\nu$ into a determinant over the 0 and 4 indices. In particular, for any pair of vectors $t$ and $s$ and any invertible matrix $O$, we have
\begin{equation}
    \det^{(4)} \left( -t_\mathtt{m} s_\mathtt{n} + \frac{1}{c^2} O_\mathtt{mn} \right) =\det^{(4)} (c^{-2} O_\mathtt{mn}) \det^{(1)} \left( -c^2 s_\mathtt{m} O^\mathtt{mn} t_\mathtt{n} + 1 \right) \approx c^{-6} \det^{(4)} (O_\mathtt{mn}) \det^{(1)} \left( -s_\mathtt{m}  O^\mathtt{mn} t_\mathtt{n}\right)
\end{equation}
which we can fuse into a single higher-dimensional determinant (Sylvester),
\begin{equation}
    \det^{(4)} (O_\mathtt{mn}) \det^{(1)} \left( -s_\mathtt{m}  O^\mathtt{mn} t_\mathtt{n}\right)= \det^{(5)}  \begin{pmatrix}
        0 & s_\mathtt{n} \\
        t_\mathtt{m}  & O_\mathtt{mn} 
    \end{pmatrix} \, , \label{seconddeterminandtransformation}
\end{equation}
where we have explicitly indicated the dimension of the determinants to clarify the formulas. We absorb the $c^{-6}$ factor inside the dilaton by redefining it as $\Phi \rightarrow \Phi - 3\log c$, so we can safely perform the $c \rightarrow \infty$ limit of the action and obtain
\begin{equation}
    \mathcal{S}_{\text{NR DBI}} = -T_4 \int{\dd \sigma^4 \Tr_S \left[ e^{-\Phi} \sqrt{ -\det  \begin{pmatrix}
        0 & \tau_\nu \partial_\mathtt{n} X^\nu \\
        \bar{\tau}_\mu \partial_\mathtt{m} X^\mu & h_{\mu \nu} \partial_\mathtt{m} X^\mu \partial_\mathtt{n} X^\nu + 2\pi \alpha' F_\mathtt{mn}  +q_\mathtt{mn}
    \end{pmatrix} }  \right] } \, . \label{NRDBI}
\end{equation}

Let us move now to the decoupling limit, $\alpha ' \rightarrow 0$. We start by setting the dilaton to $(e^{-\Phi} T_4)^{-1}=(2\pi)^3 g_s \alpha^{\prime 2}$, imposing the static gauge on the longitudinal directions to the D3 branes, $X^\mathtt{m}=\sigma^\mathtt{m}$, and rescaling the transverse directions as 
\begin{equation}
    X^4=2\pi \alpha ' \zeta \, , \qquad X^{i'}=2\pi \alpha ' \phi^{i'} \, ,
\end{equation}
which we will interpret as scalar fields.

The next step is to expand the determinant in (\ref{NRDBI}) for small values of $\alpha '$. After some algebra and using the fact that $F_\mathtt{mn}$ has vanishing trace, we get
\begin{equation}
\label{GED_action}
    \mathcal{L}_{\text{GED}}= -\frac{1}{2\pi g_s} \sum_{I=1}^{N^2}\left( \frac{1}{2} \sum_{i'=1}^5 (\partial_\mathtt{a}\phi^{i' I})^2 + \frac{1}{4} (F^I_\mathtt{ab})^2 - F^I_\mathtt{at} \partial_\mathtt{a} \zeta^I -\frac{1}{2} (\partial_\mathtt{t} \zeta^I)^2 \right) \, .
\end{equation}
The action we have obtained contains 5 non-dynamical free scalars and an off-shell action that describes the non-relativistic limit of Maxwell's equations plus a free scalar field. The latter contribution is known in the literature as ``Galilean Electrodynamics'' (GED) \cite{Santos:2004pq}.

\paragraph{Decoupling first, non-relativistic second}

To retrieve the $\mathcal{N}=4$ SYM action from (\ref{improvedDBI}), we first set the dilation to $(e^{-\Phi} T_4)^{-1}=(2\pi)^3 g_s \alpha^{\prime 2} c^{-3}$ and fix our metric $g$ to be the flat Minkowski metric in 10 dimensions. As we will be interested in the non-relativistic limit of this action, we will also set all the components of $B_2$ to zero except for $B_{04}$.

In contrast with the usual approach to this computation, here we will not impose the static gauge $X^\mathtt{m}=\sigma^\mathtt{m}$ yet. This is because later we want to perform a non-relativistic limit on the target space without touching the relativistic structure of the world-volume. Therefore, we will impose instead the  weaker condition that each coordinate longitudinal to the D3-brane depends on a different world-volume coordinate, i.e. $X^\mathtt{0}=X^\mathtt{0} (\mathtt{t})$, $X^\mathtt{1}=X^\mathtt{1} (\sigma^\mathtt{1})$, $\dots$ We identify the remaining transverse coordinates with scalar fields as follows
\begin{equation}
    X^4=2\pi \alpha ' \zeta \, , \qquad X^{i'}=2\pi \alpha ' \phi^{i'} \, .
\end{equation}

We are now allowed to compute the decoupling limit $\alpha '\rightarrow 0$. To that end, we will use the following expansion of the determinants
\begin{equation}
    \det (\mathbb{I} + \alpha ' M ) \approx 1+ \alpha ' \Tr (M) + \alpha^{\prime 2} \frac{\Tr (M)^2 - \Tr (M^2)}{2} + \mathcal{O} (\alpha^{\prime 3}) \, . 
\end{equation}
After some algebra, we obtain\footnote{Before taking the $\alpha '\rightarrow 0$ we have subtracted a contribution of the form $\frac{-c^3 \prod_{\mathtt{m}=0}^3 \partial_\mathtt{m} X^\mathtt{m}}{(2\pi)^3 g_s \alpha^{\prime 2}}$, which becomes an irrelevant constant term once static gauge is imposed.}
\begin{align}
    &\mathcal{L}_{\text{SYM}}=-\frac{c^3\prod_{\mathtt{m}=0}^3 \partial_\mathtt{m} X^\mathtt{m}}{2\pi g_s} \sum_{I=1}^{N^2} \left( \frac{1}{2} \sum_{i'=1}^5 \left[\sum_{\mathtt{a}=1}^3 \frac{(D_\mathtt{a} \phi^{i' I})^2}{(\partial_\mathtt{a} X^\mathtt{a})^2} - \frac{(D_\mathtt{t} \phi^{i' I})^2}{(\partial_\mathtt{t} X^\mathtt{t})^2}  \right]   \right.  \notag \\
    &+\frac{1}{2} \left[\sum_{\mathtt{a}=1}^3 \frac{(D_\mathtt{a} \zeta^I)^2}{(\partial_\mathtt{a} X^\mathtt{a})^2} - \frac{(D_\mathtt{t} \zeta^I)^2}{(\partial_\mathtt{t} X^\mathtt{t})^2} \right]+ \frac{(B_{40} D_\mathtt{a} \zeta^I (\partial_\mathtt{t} X^\mathtt{t}) + F_\mathtt{at})^2}{2(\partial_\mathtt{t} X^\mathtt{t}\partial_\mathtt{a} X^\mathtt{a})^2} +  \frac{1}{4} \frac{F_\mathtt{ab} F^\mathtt{ab}}{(\partial_\mathtt{a} X^\mathtt{a} \partial_\mathtt{b} X^\mathtt{b})^2} \notag \\
    &\left. +\sum_{J,K=1}^{N^2} \left[2i B_{40} f^{JK}\null_I \zeta^J\phi^{i' K} \frac{D_\mathtt{t} \phi^{i' I}}{\partial_\mathtt{t} X^\mathtt{t}}  - \sum_{i'=1}^5 \left( f^{JK}\null_I \zeta^J \phi^{i' K} \right)^2 - \sum_{\substack{i',j'=1\\i'< j'}}^5 \left( f^{JK}\null_I \phi^{i' J} \phi^{j' K} \right)^2 \right]  \right) \, , \label{N4SYM}
\end{align}
where $f^{JK}\null_I$ are the structure constants of $U(N)$. As we have chosen them to be purely imaginary, the action is real. Furthermore, if we set $B_{40}=0$ and $(\partial_\mathtt{m} X^\mathtt{m})=1$, we recover the action of $\mathcal{N}=4$ SYM.

Now we have to compute the non-relativistic limit in the same fashion as (\ref{NR_rescaling_gauge_1}) and (\ref{NR_rescaling_gauge_2}). In particular, we perform the rescalings
\begin{equation}
    X^\mathtt{a} \rightarrow \frac{X^\mathtt{a}}{c} \, , \qquad \phi^{i'} \rightarrow \frac{\phi^{i'}}{c} \, , \qquad A_{\mathtt{t}} \rightarrow \frac{A_\mathtt{t}}{c^2}  \, , \qquad A_\mathtt{a} \rightarrow \frac{A_\mathtt{a}}{c^2} \, , \label{NR_rescaling_gauge}
\end{equation}
set the static gauge and compute the $c\rightarrow \infty$ limit. The non-relativistic $\mathcal{N}=4$ SYM Lagrangian that we get has a divergent component,
\begin{equation}
    \mathcal{L}_{\text{div.}}=-\frac{c^2}{4\pi g_s} (1-B_{40}^2 ) \sum_{I=1}^{N^2} \partial_\mathtt{a} \zeta^I  \partial^\mathtt{a} \zeta^I  \, ,
\end{equation}
which can be eliminated if we set the Kalb-Ramond field to $B_{40}=\pm 1$. Once we have taken care of it, the limit is finite and gives us
\begin{equation}
\label{GED_action_2}
    \mathcal{L}_{\text{GED}}= -\frac{1}{2\pi g_s} \sum_{I=1}^{N^2}\left( \frac{1}{2} \sum_{i'=1}^5 (\partial_\mathtt{a}\phi^{i' I})^2 + \frac{1}{4} (F^I_\mathtt{ab})^2 \mp F^I_\mathtt{at} \partial_\mathtt{a} \zeta^I -\frac{1}{2} (\partial_\mathtt{t} \zeta^I)^2 \right) \, .
\end{equation}
The $\pm$ sign depends on our choice of $B$ field and can be absorbed into a redefinition of $\zeta^I$. We can see that it perfectly matches (\ref{GED_action}), as our choice (\ref{B_field_gauge}) corresponds to setting $B_{40}= 1$. Thus, the decoupling and the non-relativistic limits commute also in this regime.

\subsubsection{An alternative non-relativistic limit}
\label{sec:alternative}

In the previous section, we performed the non-relativistic limit of the $\mathcal{N}=4$ SYM action by taking the rescaling (\ref{NR_rescaling_gauge}) and computing the $c\rightarrow \infty$ limit. However, there exists an alternative method to obtain the non-relativistic limit of a relativistic action in flat space. This method involves performing instead the rescaling
\begin{equation}
    X^0 \rightarrow c X^0 \, , \qquad X^4 \rightarrow c X^4 \, ,
\end{equation}
and then computing the $c\rightarrow \infty$ limit. Interestingly, this limit produces a different result than the one we have considered. In particular, we find that the action has a divergent contribution of the form
\begin{equation}
    \mathcal{L}_{\text{div.}}=-\frac{c^2}{4\pi g_s} (1-B_{40}^2 ) \sum_{I=1}^{N^2} \left[ D_\mathtt{a} \zeta^I D^\mathtt{a} \zeta^I - \sum_{J,K=1}^{N^2}\sum_{i'=1}^5 \left( f^{JK}\null_I \zeta^J \phi^{i' K} \right)^2\right] \, ,
\end{equation}
which can still be eliminated by appropriately choosing the Kalb-Ramond field, plus a finite contribution given by
\begin{align}\label{GYM}
    &\mathcal{L}_{\text{GYM}} = -\frac{1}{2\pi g_s } \sum_{I=1}^{N^2} \left( \frac{1}{2} \sum_{i'=1}^5 \bigg[ (D_\mathtt{a}\phi^{i' I})^2 \mp \sum_{J,K=1}^{N^2} 2i f^{JK}\null_I \zeta^J \phi^{i' K} D_\mathtt{t} \phi^{i' I}\bigg]  \right. \notag \\
    &\left. - \sum_{J,K=1}^{N^2}\sum_{i'<j'} \left( f^{JK}\null_I \phi^{i' J} \phi^{j' K} \right)^2 + \frac{1}{4} (F_\mathtt{ab})^2 \pm F_\mathtt{at} D_\mathtt{a} \zeta^I -(D_\mathtt{t} \zeta^I)^2 %
    \vphantom{\left( \frac{1}{2} \sum_{i=1}^5 \bigg[ \right)} \right) \, ,
\end{align}
where, again, the $\pm$ signs depend on the sign of $B_{40}$ and can be reabsorbed into $\zeta^I$. The action we obtained is the bosonic sector of the four-dimensional Galilean Yang-Mills, first found in \cite{Bagchi:2015qcw}, and also via null reduction in \cite{Bagchi:2022twx}. In addition, its supersymmetric completion can also be obtained from the dimensional reduction of the an-isotropic non-relativistic limit of Yang-Mills computed in appendix E of \cite{Bergshoeff:2021tfn}.

At this stage, we can swap the order of limits and perform this alternative non-relativistic limit first and the decoupling limit second. It is straightforward to check that the two limits commute again, and the final result is exactly given by the action (\ref{GYM}).

The most striking difference between (\ref{GED_action_2}) and (\ref{GYM}) is that, in the former, the gauge group becomes abelian, while in the latter the non-abelian structure, both for the scalars and the field strength $F_\mathtt{mn}$, is preserved. It is immediate to see that if we perform the In\"on\"u-Wigner contraction that takes $U(N)$ to $U(1)^{N^2}$, then the action (\ref{GYM}) boils down to (\ref{GED_action_2}).

In the following section we will discuss in detail the symmetries of both the theory that we obtained from the gravity perspective and from the gauge theory one, but it is already clear that the flattening of the sphere that we see from the gravity perspective points us to pick the ``abelian'' limit as the one we are interested in. It would be interesting to explore if there exists a non-relativistic limit in the gravity side that does not flatten out the sphere and, thus, can be related to the ``non-abelian'' limit of $\mathcal{N}=4$ SYM in (\ref{GYM}).

\section{A first test: matching global symmetries}
\label{sec:global_symm}

In this section, we provide the first test of the holographic duality suggested in the previous section. First, we compute the global symmetries of non-relativistic string theory in SNC AdS$_5\times$S$^5$ by solving systematically the non-relativistic Killing equations. Second, we show that the global symmetries of the non-relativistic string action have a holographic realisation on GED with 5 uncharged massless free scalars.   

\subsection{Global symmetries of the non-relativistic string action}
\label{sec:symm_string}

Our aim is to determine all global symmetries associated with the bosonic sector of Type IIB non-relativistic string theory on the target space SNC AdS$_5\times$S$^5$. We approach this problem systematically by analysing the non-relativistic Killing equations.

Our starting point is the string action with non-relativistic target space and relativistic world-sheet. In the Nambu-Goto formulation, the action is
\begin{equation}
\label{NR_action}
    S = -\frac{T}{2} \int \dd^2 \sigma \left( \sqrt{-\tau} \tau^{\alpha\beta} h_{\alpha\beta} + \varepsilon^{\alpha\beta} m_{\alpha\beta} \right) \, , 
\end{equation}
where $\tau_{\alpha\beta}$, $h_{\alpha\beta}$ and $m_{\alpha\beta}$ are the pull-back of the longitudinal metric $\tau_{\mu\nu}$, the transverse tensor $h_{\mu\nu}$ and the non-relativistic Kalb-Ramond 2-form $m_{\mu\nu}$. The longitudinal metric and transverse tensor are written in terms of SNC vielbeine $\tau_{\mu}{}^A, e_{\mu}{}^{\hat{a}}$ as 
\begin{equation}
    \tau_{\mu\nu} = \tau_{\mu}{}^A \tau_{\nu}{}^B \eta_{AB} \, , \qquad
    h_{\mu\nu} = e_{\mu}{}^{\hat{a}} e_{\nu}{}^{\hat{b}} \delta_{\hat{a}\hat{b}} \, ,
\end{equation}
where $A$ and $\hat{a}$ are longitudinal and transverse flat indices, respectively.

All global symmetries of the action (\ref{NR_action}) generated  by the infinitesimal coordinate transformation in target space $\delta X^{\mu} = \xi^{\mu}$ are given by solving the following SNC Killing equations \cite{Bidussi:2023rfs}:
\begin{subequations}\label{SNC_Killing}
\begin{align}
\label{Lie_tau}
(\mathsterling_{\xi} \,\tau )_{\mu\nu}&= 2 \omega \tau_{\mu\nu} \, , \\
\label{Lie_h}
(\mathsterling_{\xi}\, h )_{\mu\nu} &= \Sigma_{A \hat{a}} (\tau_{\mu}{}^A e_{\nu}{}^{\hat{a}} + \tau_{\nu}{}^A e_{\mu}{}^{\hat{a}} ) \, , \\
\label{Lie_m}
(\mathsterling_{\xi}\, m )_{\mu\nu} &= \varepsilon_{AB} \Sigma^B{}_{\hat{a}} (\tau_{\mu}{}^A e_{\nu}{}^{\hat{a}} - \tau_{\nu}{}^A e_{\mu}{}^{\hat{a}} ) - \partial_{\mu} \Sigma_{\nu} +  \partial_{\nu} \Sigma_{\mu} \, ,
\end{align}
\end{subequations}
where $\Sigma_{A\hat{a}}, \Sigma_{\mu}, \omega$ are transformation parameters, functions of the spacetime coordinates.   
Notice that if we were studying relativistic strings propagating in a Lorentzian target space, one would just have to solve the equation $(\mathsterling_{\xi} \, g )_{\mu\nu} = 0$. Instead, we notice that the r.h.s. of (\ref{SNC_Killing}) is non-zero. This is due to the fact that the tensors $\tau_{\mu\nu}, h_{\mu\nu}, m_{\mu\nu}$ form an equivalence class defined up to local SNC transformations. Therefore, one has to demand that their Lie derivatives vanish up to local SNC transformations.  

After having discussed how to implement the SNC Killing equations, we are now ready to apply them to the background of our interest, namely SNC AdS$_5\times$S$^5$. The tensors $\tau_{\mu\nu}, h_{\mu\nu}, m_{\mu\nu}$ are defined following the conventions of \cite{Bidussi:2021ujm} and, for this geometry, they can be derived by the limit procedure from the Lorentzian AdS$_5\times$S$^5$ geometry in (\ref{AdS5xS5_metric}). They take the form
\begin{subequations}
\begin{align}
    \tau_{\mu\nu} \dd X^{\mu} \dd X^{\nu} &= \frac{1}{z^2} \left( -\dd t^2 + \dd z^2 \right) \, , \\
    h_{\mu\nu} \dd X^{\mu} \dd X^{\nu} &= \frac{1}{z^2} \dd x^i \dd x^j \delta_{ij} +  \dd x^{i'} \dd x^{j'} \delta_{i' j'} \, ,  \\
    m_{\mu\nu} &= 0 
 \end{align}
\end{subequations}
for a unit radius. 
The SNC vielbeine associated with these tensors are given by
\begin{subequations}
\begin{align}
    \tau_{\mu}{}^A &= z^{-1} \delta_{\mu}^A \, ,& 
    A &= 0, 4 \, , & \text{longitudinal AdS$_5$}\\
    e_{\mu}{}^a &= z^{-1} \delta_{\mu}^a \, ,& 
    a &= 1,2,3 \, , & \text{transverse AdS$_5$} \\
    e_{\mu}{}^{a'} &= \delta_{\mu}^{a'} \, , & 
    a' &= 5, ...., 9 \, , &  \text{transverse S$^5$}
 \end{align}
\end{subequations}
where we have split the flat spatial indices as $\hat{a} = (a, a')$. In this geometry, the set of SNC Killing equations (\ref{SNC_Killing}) reads as follows.\footnote{In what follows, the index of the Killing vector $\xi$ is raised and lowered with the flat metrics $\eta_{AB}$ and $\delta_{ab}$.} 

\noindent The $\tau$ equation (\ref{Lie_tau}) reads as
\begin{subequations}\label{tau_Killing}
\begin{align}
\label{tau_aA}
    \partial_{\hat{a}} \xi_A &= 0 \, , \\
\label{tau_AB}   
    \partial_A \xi_B + \partial_B \xi_A &= 2 \left( \omega + \frac{1}{z} \xi_4 \right) \eta_{AB} \, ,
    \end{align}
\end{subequations}
The $h$ equation (\ref{Lie_h}) reads as
\begin{subequations}\label{h_Killing}
\begin{align}
\label{h_Aa}
    \partial_A \xi_a &= \Sigma_{Aa} \, , \\
\label{h_Aa'}
    \partial_A \xi_{a'} &= \frac{1}{z} \Sigma_{Aa'} \, , \\
\label{h_aa'}
    \partial_a \xi_{a'} + \frac{1}{z^2} \partial_{a'} \xi_a &= 0 \, , \\
\label{h_ab}
    \partial_a \xi_b + \partial_b \xi_a &= \frac{2}{z} \xi_4 \delta_{ab} \, , \\
\label{h_a'b'}
    \partial_{a'} \xi_{b'} + \partial_{b'} \xi_{a'} &= 0 \, .
    \end{align}
\end{subequations}
The $m$ equation (\ref{Lie_m}) reads as
\begin{subequations}\label{m_Killing}
\begin{align}
\label{m_AB}
    \partial_A \Sigma_B - \partial_B \Sigma_A &= 0 \, , \\
\label{m_Aa}
    \partial_A \Sigma_a - \partial_a \Sigma_A &= \frac{1}{z^2} \varepsilon_{AB} \Sigma^B{}_a \, , \\
\label{m_Aa'}
    \partial_A \Sigma_{a'} - \partial_{a'} \Sigma_A &= \frac{1}{z} \varepsilon_{AB} \Sigma^B{}_{a'} \, , \\
\label{m_ab_hatted}
    \partial_{\hat{a}} \Sigma_{\hat{b}} - \partial_{\hat{b}} \Sigma_{\hat{a}} &= 0 \, .
    \end{align}
\end{subequations}
First, we notice that we can locally solve equations (\ref{m_AB}) and (\ref{m_ab_hatted}) in terms of generic functions of the spacetime coordinates, $f$ and $g$, as follows: 
\begin{eqnarray}
    \Sigma_A = \partial_A f \, , \qquad\qquad
    \Sigma_{\hat{a}} = \partial_{\hat{a}} g \, .
\end{eqnarray}
Then, we solve equations (\ref{m_Aa}) and (\ref{m_Aa'}) in terms of these functions, 
\begin{eqnarray}
\label{Sigma_Aa}
    \Sigma_{Aa} &=& z^2 \varepsilon_A{}^B \left( \partial_B \partial_{a} g - \partial_a \partial_B f \right) \equiv z^2 \, \varepsilon_A{}^B \partial_B \partial_a \mathfrak{F} \, , \\
\label{Sigma_Aa'}
    \Sigma_{Aa'} &=& z \, \varepsilon_A{}^B \left( \partial_B \partial_{a'} g - \partial_{a'} \partial_B f \right) \equiv z \, \varepsilon_A{}^B \partial_B \partial_{a'} \mathfrak{F} \, , 
\end{eqnarray}
where $\mathfrak{F} \equiv g - f$. 

Now we move to solving the $\tau$ SNC Killing equations (\ref{tau_Killing}). From (\ref{tau_aA}), we get that $\xi_A$ is independent of the transverse coordinates $x^{\hat{a}}$. To solve (\ref{tau_AB}), it is convenient to introduce the light-cone basis for longitudinal coordinates, namely
\begin{eqnarray}
    x^{\pm} = X^0 \pm X^4 = t \pm z \, , 
\end{eqnarray}
so equation (\ref{tau_AB}) becomes
\begin{eqnarray}
    \partial_{\pm} \xi_{\pm} = 0 \, , \qquad\qquad
    \partial_+ \xi_- + \partial_- \xi_+ = - \omega - \frac{1}{z}\, (\xi_+ - \xi_-) \, ,
\end{eqnarray}
whose most generic solution is
\begin{eqnarray}
    \xi^+ =  \xi^+ (x^+) \, , \qquad\quad
    \xi^- =  \xi^- (x^-)  \, .
\end{eqnarray}
Next, we solve the $h$ equations (\ref{h_Killing}). The most general solution to equation (\ref{h_a'b'}) is
\begin{eqnarray}
\label{most_gen_1}
  \xi^{a'} \partial_{a'} = k^{a'}(x^+, x^-, x^a) \partial_{a'} + \omega^{a'b'}(x^+, x^-, x^a)(x^{a'} \partial_{b'} - x^{b'} \partial_{a'} ) \, ,  
\end{eqnarray}
whereas the most general solution to equation (\ref{h_ab}) is
\begin{eqnarray}
\label{most_gen_2}
  \xi^{a} \partial_{a} = \left[ k^{a}(x^+, x^-, x^{a'}) + \frac{x^a}{z} \xi_4 \right] \partial_a + \omega^{ab}(x^+, x^-, x^{a'})(x^a \partial_b - x^b \partial_a ) \, ,  
\end{eqnarray}
where $\xi_4 = \xi^+ - \xi^-$. The strategy now is to find how the remaining Killing vector equations further constrain the coefficients $k^{\hat{a}}, \omega^{\hat{a}\hat{b}}$ appearing in (\ref{most_gen_1}) and (\ref{most_gen_2}). 
First, we start with equation (\ref{h_aa'}). By applying $\partial_b$ to (\ref{h_aa'}) and anti-symmetrising in $a, b$, (and by repeating the same procedure on the primed directions) we get
\begin{eqnarray}
    \partial_{a'} \partial_{[a} \xi_{b]} &=& 0 \, , \qquad\Longrightarrow\qquad
    \omega_{ab} = \omega_{ab}(x^+, x^-) \, , \\
        \partial_{a} \partial_{[a'} \xi_{b']} &=& 0 \, , \qquad\Longrightarrow\qquad
    \omega_{a'b'} = \omega_{a'b'}(x^+, x^-) \, . 
\end{eqnarray}
Then, equation (\ref{h_aa'}) imposes the following relation between the coefficients $k^{\hat{a}}$:
\begin{eqnarray}
    z^2 \partial_a k_{a'} + \partial_{a'} k_a = 0 \, , 
\end{eqnarray}
which implies 
\begin{eqnarray}
    k_a &=& u_a (x^+, x^-) - z^2 v_{a'a}(x^+, x^-) \, x^{a'} \, , \\
    k_{a'} &=& u_{a'} (x^+, x^-) + v_{a'a}(x^+, x^-) \, x^a \, .
\end{eqnarray}
Next, we consider equation (\ref{h_Aa}) [resp. (\ref{h_Aa'})] combined with (\ref{Sigma_Aa}) [resp. (\ref{Sigma_Aa'})]. By differentiating with respect to $x^b$ (resp. $x^{b'}$) and by taking the anti-symmetrisation on $a,b$ (resp. $a', b'$), we get
\begin{eqnarray}
    \partial_A  \partial_{[a} \xi_{b]} &=& 0 \, , \qquad \Longrightarrow\qquad
    \omega_{ab} = \text{const.} \, ,   \\
    \partial_A  \partial_{[a'} \xi_{b']} &=& 0 \, , \qquad \Longrightarrow\qquad
    \omega_{a'b'} = \text{const.}  
\end{eqnarray}
Then, by combining and deriving (\ref{h_Aa}) and (\ref{h_Aa'}), we obtain the following condition:
\begin{equation}
    \partial_A \partial_{a'} \xi_a - z^2 \partial_A \partial_a \xi_{a'} = 0 \, , \qquad\Longrightarrow\qquad
    \partial_A (z^2 v_{a'a}) + z^2 \partial_A v_{a'a} = 0 \, .
\end{equation}
For $A=0$, the above condition gives $v_{a'a} = v_{a'a} (z)$. For $A=1$, it gives the following differential equation which can be immediately solved: 
\begin{eqnarray}
\partial_z v_{a'a} = - \frac{v_{a'a}}{z}  \, , \qquad \Longrightarrow \qquad
v_{a'a} = \frac{\theta_{a'a}}{z} \, , \qquad \theta_{a'a} = \text{const.}  
\end{eqnarray}
At this stage, one needs to check that (\ref{h_Aa}) and (\ref{h_Aa'}) are separately satisfied, i.e., that we can define a consistent $\mathfrak{F}$. From (\ref{h_Aa'}) we get
\begin{eqnarray}
\label{check}
    \partial_A \left(\frac{\theta_{a'a}}{z} x^a \right) = \varepsilon_A{}^B \partial_B \partial_{a'} \mathfrak{F} \qquad\Longrightarrow \qquad\left\{ \begin{array}{c}
       0= \partial_z \partial_{a'} \mathfrak{F}  \\
       -\cfrac{\theta_{a'a}}{z^2} x^a  = \partial_t \partial_{a'} \mathfrak{F}
    \end{array}\right. \ .  
\end{eqnarray}
By taking the derivative with respect to $z$ of the second equation in (\ref{check}), commuting the derivatives $\partial_z$ and $\partial_t$, and by substituting the first equation on it, we find that the only possible solution is $\theta_{a'a}=0$.

At this stage, all coefficients have been maximally restricted, except for $u_{\hat{a}}$. To proceed further, we first consider equation (\ref{h_Aa'}) combined with $(\ref{Sigma_Aa'})$. By hitting this equation with $\eta^{AB}\partial_B$ we get the wave equation 
\begin{eqnarray}
    (\partial_t^2 - \partial_{z}^2) \, \xi_{a'} = 0 \, , \qquad\Longrightarrow\qquad
    u_{a'} = u_{a'}^+ (x^+) + u_{a'}^- (x^-) \, , 
\end{eqnarray}
namely $u_{a'}$ is a separable function of $x^+$ and $x^-$, which we expand in the Laurent series as 
\begin{eqnarray}
    u_{a'}^+ (x^+) = \sum_{n\in \mathbb{Z}} c^n_{a'} (t+z)^n  \, , \qquad
    u_{a'}^- (x^-) = \sum_{n\in \mathbb{Z}} d^n_{a'} (t-z)^n \, , 
\end{eqnarray}
where $c^n_{a'}$ and $d^n_{a'}$ are constants.
Finally, we consider (\ref{h_Aa}) combined with (\ref{Sigma_Aa}), and we hit it with $\eta^{AB}\partial_B$ as before. In this case, we get 
\begin{eqnarray}
\label{modified_wave_equation}
    \left( \partial_t^2 - \partial_z^2 + \frac{2}{z} \partial_z \right) \xi_a = 0 \, .
\end{eqnarray}
The most generic meromorphic solution to equation (\ref{modified_wave_equation}) is of the form 
\begin{eqnarray}
    \xi_a=\sum_{n\in \mathbb{Z}} a^n_a (t-z)^n (t+nz) + b^n_a (t+z)^n (t-nz) \ ,
\end{eqnarray}
where $a^n_a$ and $b^n_a$ are constants. The vector field $\xi^a \partial_a$ contains a term proportional to $\xi_4$, and therefore equation (\ref{modified_wave_equation}) imposes a further restriction on $\xi_4$. Moreover, equation (\ref{tau_AB}) relates $\xi_4$ with $\xi_0$, and hence $\xi_0$ will also be restricted. By doing that, one gets that the most generic $\xi_4$ and $\xi_0$ are 
\begin{eqnarray}
    \xi_4 &=& \alpha z + \beta z t \, , \\
    \xi_0 &=& \gamma - \alpha t - \frac{\beta}{2} (t^2 + z^2) \, ,
\end{eqnarray}
where $\alpha, \beta, \gamma$ are constants. Notice that they still are in the separable form in terms of the light-cone variables $x^{\pm}$, as expected.

At this stage, we have maximally restricted the most generic Killing vector $\xi^{\mu}$ that satisfies the SNC Killing equations. The most generic Killing vector $\xi^{\mu}$ contains a family of Killing vectors, parametrised by the free constant coefficients $\alpha, \beta, \gamma, a^n_a, b^n_a, c^n_{a'}, d^n_{a'}, \omega_{ab}, \omega_{a'b'}$.  
In summary, the Killing vectors admitted by the action (\ref{NR_action}) in SNC AdS$_5\times$S$^5$ are
\begin{subequations}\label{KV}
\begin{align}
    H &= \partial_t \, , \\
    D &= t\, \partial_t + z \, \partial_z + x^a \partial_a \, , \\
    K &= (t^2 + z^2) \partial_t +2 \,z\, t\, \partial_z +2 \, t \, x^a \partial_a\, , \\
    P^{(n)}_a &= (t-z)^n (t+nz) \partial_a \, , \qquad P^{(n)}_{a'} =  (t-z)^n  \partial_{a'} \, ,  \\
    \tilde{P}^{(n)}_a &= (t+z)^n (t-nz) \partial_a \, , \qquad \tilde{P}^{(n)}_{a'} =  (t+z)^n\partial_{a'} \, ,\\
    J_{ab} &= x^a \partial_b - x^b \partial_a \, , \hspace{2cm}   J_{a'b'} = x^{a'} \partial_{b'} - x^{b'} \partial_{a'} \, .
\end{align}
\end{subequations}
The non-vanishing commutation relations between the Killing vectors (\ref{KV}) are 
\begin{subequations} \label{global_symm_bulk}
	\begin{align}
    [H, D] &= H \, , \qquad\qquad
 [H, K] = 2 D \, , & \qquad
 [K, D] &= - K \, , \\
\null [P_{a}^{(n)}, D] &= - n P_{a}^{(n)}\, ,& \qquad
\null [\tilde{P}_{a}^{(n)}, D] &= - n \tilde{P}_{a}^{(n)} \, , \\
\null [P_{a'}^{(n)}, D] &= - n P_{a'}^{(n)} \, ,& \qquad
\null [\tilde{P}_{a'}^{(n)}, D] &= - n \tilde{P}_{a'}^{(n)} \, , \\
\null [P_{a}^{(n)}, H ] &= - (n+1) P_{a}^{(n-1)} \, , &\qquad
\null [\tilde{P}_{a}^{(n)}, H ] &= - (n+1) \tilde{P}_{a}^{(n-1)} \, , \\
\null [P_{a'}^{(n)}, H ] &= - n P_{a'}^{(n-1)} \, , & \qquad
\null [\tilde{P}_{a'}^{(n)}, H ] &= - n \tilde{P}_{a'}^{(n-1)} \, , \\
\null [P_{a}^{(n)}, K] &= - (n-1) P_{a}^{(n+1)} \, ,& \qquad
\null [\tilde{P}_{a}^{(n)}, K] &= - (n-1) \tilde{P}_{a}^{(n+1)} \, , \\
\null [P_{a'}^{(n)}, K] &= -n P_{a'}^{(n+1)} \, , &\qquad
\null [\tilde{P}_{a'}^{(n)}, K] &= -n \tilde{P}_{a'}^{(n+1)} \, , \\
\null [J_{ab}, P_c^{(n)}] &= \delta_{bc} P_a^{(n)}- \delta_{ac} P_b^{(n)} \, , &
\qquad
\null [J_{ab}, \tilde{P}_c^{(n)}] &= \delta_{bc} \tilde{P}_a^{(n)} - \delta_{ac} \tilde{P}_b^{(n)} \, , 
\\
\null [J_{a'b'}, P_{c'}^{(n)}] &= \delta_{b'c'} P_{a'}^{(n)} - \delta_{a'c'} P_{b'}^{(n)} \, , &
\qquad
\null [J_{a'b'}, \tilde{P}_{c'}^{(n)}] &=
\delta_{b'c'} \tilde{P}_{a'}^{(n)} - \delta_{a'c'} \tilde{P}_{b'}^{(n)} \, , 
\\
\null [J_{ab}, J_{cd}] &= 2 \delta_{c[b} J_{a]d} - 2 \delta_{d[b} J_{a]c} \, , &
\qquad
\null [J_{a'b'}, J_{c'd'}] &= 2 \delta_{c'[b'} J_{a']d'} - 2 \delta_{d'[b'} J_{a']c'} \, . 
	\end{align}
\end{subequations}
The isometry algebra we found is infinite dimensional. The generators $H, D, K$ form an $\mathfrak{sl}(2,\mathbb{R})$ finite subalgebra, and they describe the isometries of AdS$_2$ appearing in the $\tau_{\mu\nu}$ metric. The infinite tower of translations in the transverse coordinates, which depend on both $t$ and $z$, are isometries that cannot be found by taking the standard Newton-Hooke contraction of the isometry algebra of AdS$_5\times$S$^5$ \cite{Bagchi:2009my, Bergshoeff:2023fil}. However, as we shall see in section \ref{sec:symm_gauge}, they have an important holographic realisation.

\subsubsection{Symmetries of the full non-relativistic supergravity solution}

The world-sheet action (\ref{NR_action}) has target space SNC AdS$_5\times$S$^5$, and in order to be free of the Weyl anomaly, it needs to satisfy the non-relativistic type IIB supergravity equations of motion. Because of that, any symmetry of the metric tensors also needs to be a symmetry of the fluxes, of the dilaton and of the 5-form Lagrange multiplier, since they are related by the non-relativistic Einstein equations obtained by setting to zero the beta function. Here, we show that the symmetries of the string world-sheet sigma model computed in the previous section are also symmetries of the full non-relativistic type IIB supergravity solution.  All the relevant expressions regarding equations of motion and symmetries of the non-relativistic type IIB supergravity can be found in \cite{Bergshoeff:2023ogz}.

\paragraph{RR forms.}
The relativistic AdS$_5\times$S$^5$ type IIB supergravity solution has vanishing RR fields, except of the 5-form field strength, which is 
\begin{align}
    F^{(5)} &\sim \frac{e^{-\Phi}}{R} \left[ \text{dvol}(\text{AdS}_5) + \text{dvol}(\text{S}^5) \right]  \\
    &\sim \frac{e^{-\Phi}}{R} \left[ \frac{R^5}{z^5} \dd t\wedge \dd z\wedge \dd x^1 \wedge\dd x^2 \wedge\dd x^3 + R^5 \sqrt{\frac{1-\frac{y^2}{4}}{\left( 1+\frac{y^2}{4} \right)^9}} \dd \phi \wedge \dd y_1 \wedge \dd y_2\wedge \dd y_3\wedge \dd y_4  \right]\, , \notag
\end{align} 
where $\Phi$ is constant. To take the non-relativistic limit, we need to rescale the coordinates and parameters as in (\ref{NR_rescaling}), and to fix the dilaton as $\Phi = \log c$. After taking $c\to \infty$, the RR 5-form becomes\footnote{The non-relativistic limit of the self-duality condition only involves the components of $F^{(5)}_{\text{\scriptsize NR\normalsize}}$ with zero or one longitudinal coordinates, and thus it is trivially fulfilled.}
\begin{eqnarray}
\label{NR_RR_5_form}
    F^{(5)}_{\text{\scriptsize NR\normalsize}} \sim R^{-1} \text{dvol}(\text{AdS}_5) \, , 
\end{eqnarray}
where the volume form of the 5-sphere dropped out in the limit, in agreement with the fact that S$^5$ flattens out and becomes the 5-dimensional Euclidean space. 

Under a local boost transformation, each RR $p$-form transforms with a term proportional to an RR $(p-2)$-form. Since in our supergravity solution all RR forms vanish except of the 5-form field strength, we only need to impose that the Lie derivative of (\ref{NR_RR_5_form}) vanishes. Happily, all Killing vectors given in (\ref{KV}) satisfy this condition.

\paragraph{Dilaton.}
The condition to impose is that the Lie derivative of $e^{\phi}$ is proportional to $e^{\phi}$. Since the non-relativistic dilaton $\phi$ vanishes, this implies that the constant of proportionality needs to vanish as well, which corresponds to the condition $\omega = 0$. Then the parameter $\omega$ only enters in the condition \eqref{Lie_tau}. This means that the Killing vectors need to fulfil \eqref{Lie_tau} with $\omega = 0$, namely their Lie derivatives of $\tau_{\mu\nu}$ has to vanish. We checked that this is actually the case for the whole set of Killing vectors given in \eqref{KV}.

\paragraph{Lagrange multiplier $\mathcal{A}^{(5)}$.} First of all, we notice that our solution fulfills the following conditions
\begin{eqnarray} \label{sol_cond}
    F^{(5)}_{\text{\scriptsize NR\normalsize}} \wedge B^{(2)}_{\text{crit.}}=0 \, , \qquad\qquad
    F^{(3)}_{\text{\scriptsize NR\normalsize}}=0 \, ,
\end{eqnarray}
where we defined $(B^{(2)}_{\text{crit.}})_{\mu\nu} = \tau_{\mu}{}^A \tau_{\nu}{}^B \varepsilon_{AB}$. Thanks to this condition, the constraint imposed by the equation of motion for $\mathcal{A}^{(5)}$ is automatically satisfied. 

The non-relativistic type IIB supergravity equations of motion impose that $\mathcal{A}^{(5)}$ has to be proportional to $F^{(5)}_{\text{\scriptsize NR\normalsize}}$. Therefore, the Lie derivative of $\mathcal{A}^{(5)}$ is zero for all Killing vectors (\ref{KV}). On the other hand, the variation of $\mathcal{A}^{(5)}$ under a local boost transformation is also zero because of the condition (\ref{sol_cond}).

We conclude that the Killing vectors (\ref{KV}) that are symmetries of the string sigma model  are also symmetries of the full non-relativistic type IIB supergravity solution.

\subsection{Symmetries of the abelian gauge theory}
\label{sec:symm_gauge}

As we have seen in section \ref{sec:gauge_persp}, there are two options for taking the non-relativistic limit in the DBI action, which give rise to an abelian or a non-abelian non-relativistic gauge theory. Here we support our claim that the abelian theory given in (\ref{GED_action}) is the holographic dual of non-relativistic string theory in SNC AdS$_5\times$S$^5$ by matching their symmetries.       

The abelian non-relativistic gauge theory (\ref{GED_action}) is described by $N^2$ copies of the same simple system given by Galilean Electrodynamics (GED) supplemented with $5$ copies of uncharged massless free scalar fields $\phi^{i'}$. The symmetries of the GED action have been found in \cite{GED_symmetries}, which in $3+1$ dimensions are given by the following set of generators:
\begin{eqnarray}
\label{GED_on_shell_symm}
\notag
    &&M^{(n)}_{\mathtt{a}} = \mathtt{t}^{n+1} \partial_{\mathtt{a}} \, , \qquad
    H = \partial_{\mathtt{t}} \, , \qquad
    D = \mathtt{t} \partial_{\mathtt{t}} + \sigma^{\mathtt{a}} \partial_{\mathtt{a}} \, , \qquad
    J_{\mathtt{a}\mathtt{b}} = \sigma^{\mathtt{a}}\partial_{\mathtt{b}} - \sigma^{\mathtt{b}} \partial_{\mathtt{a}} \, , \\
    &&K = \mathtt{t}^2 \partial_{\mathtt{t}} + 2 \mathtt{t} \sigma^{\mathtt{a}} \partial_{\mathtt{a}} \, . 
\end{eqnarray}
By comparing them with the Killing vectors obtained from the string theory side, we immediately see that $H, D, K, M^{(n)}_{\mathtt{a}}, J_{\mathtt{a}\mathtt{b}}$ of (\ref{GED_on_shell_symm}) precisely match the expression of $H, D, K$, $P^{(n)}_a, J_{ab}$ in (\ref{KV}) evaluated on the Penrose boundary $z=0$, and up to identifying the string coordinates $t$ and $x^i$ with the coordinates $\mathtt{t}$ and $\sigma^{\mathtt{a}}$ on NC Mink$_4$ where the non-relativistic gauge theory lives. Notice that at the Penrose boundary $z=0$, the generators $\tilde{P}^{(n)}_a$ have the same expression as the $P^{(n)}_a$, and therefore we just need to consider one copy of them. The same happens for $\tilde{P}^{(n)}_{a'}$ and $P^{(n)}_{a'}$. As originally pointed out in \cite{Festuccia:2016caf}, it is interesting to note that only in $3+1$ dimensions the symmetries of GED contains the finite dimensional Galilean conformal algebra \cite{Bagchi:2009my}, where the generator $K$ has the meaning of special conformal transformation. 

At this point, from the string theory side, we note that we have extra generators $J_{a'b'}$ and $P^{(n)}_{a'}$ which we have not yet given an holographic explanation. From the gauge theory perspective, they generate an infinite dimensional non-compact R-symmetry acting on the scalar fields $\phi^{i'}$. Their realisation on the Penrose boundary in terms of holographic gauge theory coordinates is given by 
\begin{eqnarray}
    J_{i'j'} = \phi^{i'} \frac{\partial}{\partial \phi^{j'}} - \phi^{j'} \frac{\partial}{\partial \phi^{i'}} \, , \qquad\qquad
    P^{(n)}_{i'} = \mathtt{t}^n \frac{\partial}{\partial \phi^{i'}} \, . 
\end{eqnarray}
Their action on the fields $\phi^{i'}$ consists in rotations and time dependent translations. The rotation symmetries are something that we are already familiar with from the relativistic theory. However, the time dependent translations are new, and they are possible because the abelian theory (\ref{GED_action}) does not contain commutators of the type $[\phi^{i'}, \phi^{j'}]$, nor time derivatives of the scalar fields.

\vspace{5mm}

\noindent {\bf Note Added (Aug 2025):} \labeltext{`Note Added'}{NoteAdded} The proposed non-relativistic AdS/CFT correspondence was initially based on matching the on-shell symmetries of the GED theory computed in \cite{Festuccia:2016caf} with the Killing vectors of the non-relativistic string theory. Recently, the symmetries of the GED theory have been revisited in \cite{GED_symmetries}. It has been found that most of what were identified in \cite{Festuccia:2016caf} as on-shell symmetries are actually fully off-shell symmetries. Furthermore, it has been found in \cite{GED_symmetries} that the number of symmetries of the GED theory with five free scalars given in eq.~\eqref{GED_action} is larger than the number of Killing vectors found for the bulk theory. For this reason, the proposed GED theory with five free scalars cannot be the holographic dual candidate.

In contrast, the symmetries of Galilean Yang-Mills theory with five non-abelian interacting scalars, whose Lagrangian is given in eq.~\eqref{GYM} have been carefully analysed in \cite{GED_symmetries}, and they have been proven to be in a one-to-one correspondence with the Killing vectors evaluated at the Penrose boundary of the bulk supergravity solution. This contrasts with our earlier intuition that flattening the 5-sphere pointed to the Abelian theory as the holographic dual. In light of recent developments, we are thus led to identify the GYM theory with five interacting scalars as the correct holographic dual.  

This note is also published in PRD as an Erratum of this article.

\section{Conclusions}
\label{sec:conclusions}

In this article, we construct a new type of holographic correspondence by considering the non-relativistic limit of the well-known AdS$_5$/CFT$_4$ correspondence proposed by Maldacena. We claim that the duality between string theory in AdS$_5\times$S$^5$ and $\mathcal{N}=4$ Super-Yang-Mills survives in the non-relativistic limit, where it becomes a duality between a non-relativistic string theory in String Newton-Cartan AdS$_5\times$S$^5$ and Galilean Electrodynamics in Newton-Cartan 3+1 dimensional Minkowski spacetime, supplemented with five uncharged massless free scalars. The key observation that led us to propose this duality is the fact that the non-relativistic limit commutes with the decoupling limit, both for the gravitational regime and the gauge theory regime, together with the symmetry matching.

The next step would be to provide quantitative evidence of the duality. One of the most well-known tests of the relativistic AdS/CFT correspondence is the equality between the energies of semiclassical strings and the conformal dimensions of single-trace operators. This test can be extended to the non-relativistic setting we are considering here. On the string theory side, there have been some recent advances on computing the spectrum of classical strings and their quantisation \cite{Fontanella:2021btt, Fontanella:2023men, Fontanella:2021hcb, deLeeuw:2024uaq}. On the gauge theory side, it is known that GED is only (Galilean) conformal in 3+1 dimensions \cite{Festuccia:2016caf}. Therefore, a natural object similar to the relativistic dilatation operator should exist and computable in perturbation theory. Although this quantity has not been studied yet, Feynman diagram techniques for the GED Lagrangian may be borrowed from \cite{Banerjee:2022uqj}. Furthermore, it would be interesting to see if there exists a spin chain interpretation as the one that exists for $\mathcal{N}=4$ SYM \cite{Minahan:2002ve, Beisert:2003tq,Beisert:2004hm}.

What we proposed in this article is a holographic duality between a free gauge theory, i.e. GED with free scalars, and a string theory derived from an SNC limit of AdS$_5\times$S$^5$ first introduced in \cite{Gomis:2005pg}. This is not the first time holography is proposed between free gauge theories and gravitational ones, see e.g. \cite{Douglas:2010rc} and references therein. The non-relativistic string spectrum in light-cone gauge above the BMN-like folded vacuum was identified in \cite{deLeeuw:2024uaq}, and it was shown to be captured by massive and massless \emph{free} scalar fields in AdS$_2$. All higher corrections were shown to vanish up to sextic terms in perturbation theory. To match this result from the gauge theory side in terms of scaling dimensions of some gauge invariant operators (still to identify), we should not expect that the gauge theory produces any anomalous dimension. This is an indication that the dual gauge theory should be free, in agreement with our proposal.

When taking the non-relativistic limit on the DBI, we saw that there is a second possibility, leading to a Galilean Yang-Mills theory in 3+1 dimensions coupled to 5 charged massless scalar fields with a potential. 
The natural question is what would be its holographic dual theory. The presence of the commutators suggests that the dual theory should not show an internal symmetry with time dependent translations acting on the scalar fields. Intuitively, this might be realised by a non-relativistic limit that does not flatten out the 5-sphere in the bulk. As the flattening of the sphere is caused by the rescaling of the AdS radius, which is tied to the radius of the sphere, this hypothetical limit would require us to reach the non-relativistic SNC geometry without rescaling the radius. Potentially, this might require us to consider a 7-dimensional $\tau_{\mu \nu}$ metric that describes AdS$_2\times$S$^5$, and with a warped $\mathbb{R}^3$ transverse space.

The fact that our construction points us to $N^2$ copies of Galilean Electrodynamics has interesting implications for the dual string theory. For example, it is immediate to see that the free energy associated to this theory will always be proportional to $N^2$. Therefore, this suggests that non-relativistic string theory in String Newton-Cartan AdS$_5\times$S$^5$ does not have a Hagedorn phase transition. Testing this prediction would be a straightforward check of our proposal.\footnote{We thank Troels Harmark for pointing us towards this idea.}

In this paper, we have only analysed the bosonic sector of the duality. An obvious future task would be to include fermions in this holographic construction. We expect the supersymmetric completion of the string action in SNC AdS$_5\times$S$^5$, which was already done in \cite{Gomis:2005pg}, to be dual to the 3+1 dimensional analog of the supersymmetric GED in 2+1 dimensions constructed in \cite{Baiguera:2022cbp} via null reduction. 
From the gauge theory side, we find that one of the scalars behaves differently from the other five. This could be explained in two ways: either supersymmetry is broken, or it is a consequence of the supercharges closing into Galilean symmetry instead of Poincar\'e. We expect the latter to be the correct reasoning, as from the string theory side we know supersymmetry does not break \cite{Gomis:2005pg}.

An interesting feature of our construction is that the non-relativistic D3-brane metric still retains a notion of horizon. This gives hope that there might exist non-relativistic black holes in String Newton-Cartan geometry. Then, it would be interesting to study the holographic correspondence between non-relativistic black holes with asymptotic geometry SNC AdS$_5\times$S$^5$ and GED in 3+1 dimensions with scalar fields at finite temperature.

\section*{Acknowledgments}
%%%%%%%%%%%%%%%%%%%%%%%%%%%%%%%%%%%%%%

We thank Eric Bergshoeff, Marius de Leeuw, Sergey Frolov, Jelle Hartong and Tristan McLoughlin for useful discussions. In particular, we thank Jelle Hartong for important discussions related to the Penrose conformal boundary and Killing vectors of a String Newton-Cartan geometry.
We thank Marius de Leeuw, Rafael Hernández, Roberto Ruiz and  Arkady Tseytlin for valuable feedback on a draft of this work. 
AF is supported by the SFI and the Royal Society under the grant number RFF$\backslash$EREF$\backslash$210373. JMNG is supported by the Deutsche Forschungsgemeinschaft (DFG, German Research Foundation) under Germany's Excellence Strategy -- EXC 2121 ``Quantum Universe'' -- 390833306.
AF thanks Lia for her permanent support. 

\newpage 
\begin{appendices}

\section{Conventions}
\label{app:Conventions}

For a generic object $\mathcal{O}^A$, we define its light-cone combinations as
\begin{equation}\label{LC_comb}
\mathcal{O}^{\pm} \equiv \mathcal{O}^0 \pm \mathcal{O}^4 \ , \qquad\qquad
\mathcal{O}_{\pm} \equiv \frac{1}{2}\left( \mathcal{O}_0 \pm \mathcal{O}_4 \right) \ .
\end{equation}
The 2-dimensional Minkowski metric in light-cone coordinates has non-vanishing components $\eta_{+-} = -1/2$ and $\eta^{+-} = -2$. Our convention for the Levi-Civita tensor is $\varepsilon^{04} = - \varepsilon_{04} = + 1$, which in light-cone coordinates reads as $\varepsilon_{+-} = \frac{1}{2}$, $\varepsilon^{+-} = -2$.
Our convention for $p$-forms is $\omega_p = \frac{1}{p!} \omega_{\mu_1 \cdots \mu_p} \dd x^{\mu_1}\wedge \cdots \wedge \dd x^{\mu_p}$.
The indices used in this paper are summarised in Tables \ref{indices_table_gravity} and \ref{indices_table_gauge}.

\vspace{5mm}
\renewcommand{\arraystretch}{1.5}
\begin{table}[h!]
\begin{center}
        \begin{tabular}{|c|c|c|c|}
        \hline
       Indices  & Range & Purpose of the index\\
       \hline \hline
        $i, j, k$ & $1,2,3$ & spatial coordinates $x^i$ on D3-brane  \\
        $m, n$ & $1, ..., 4$ & coordinates $y^m$ on S$^5$  \\
        $i', j', k'$ & $5,..., 9$  & flattened out coordinates $x^{i'}$  on S$^5$  \\
        $A, B$ & $0,4$ & flat longitudinal indices \\
        $a,b$ & $1,2,3$ & flat transverse indices in AdS$_5$ \\
         $a',b'$ & $5,..., 9$ & flat transverse indices in S$_5$ \\
         $\alpha, \beta$ & $0,1$ & 
         coordinates $\sigma^{\alpha}$ of the string world-sheet\\
        \hline
    \end{tabular}
    \caption{Indices used in the gravity description. }
      \label{indices_table_gravity}
    \end{center}
\end{table}

\renewcommand{\arraystretch}{1.5}
\begin{table}[h!]
\begin{center}
        \begin{tabular}{|c|c|c|c|}
        \hline
       Indices  & Range & Purpose of the index\\
       \hline \hline
        $\mathtt{m}, \mathtt{n}$ & $0, ..., 3$ & world-volume coordinates $\sigma^{\mathtt{m}}$ of the D3-brane \\
        $\mathtt{a}, \mathtt{b}$ & $1,2,3$ & spatial coordinates $\sigma^{\mathtt{a}}$ in NC Mink$_4$  \\
        $p,q,r$ & $4,..., 9$ & transverse directions to the D3-brane \\
        $i', j', k'$ & $5,..., 9$  & R-symmetry index of the scalar fields $\phi^{i'}$ \\
        $I,J,K$ & $1,..., N^2$ & label of generators of $U(N)$ \\
        \hline
    \end{tabular}
    \caption{Indices used in the gauge description. }
      \label{indices_table_gauge}
    \end{center}
\end{table}

\end{appendices}

%%%%%%%%%%%%%%%% BIBLIOGRAPHY %%%%%%%%%%%%%%%%

\bibliographystyle{nb}

\bibliography{Biblio.bib}

\end{document}